%% file: Sindex_correction-v6.0.tex
\documentclass{aa}

\usepackage{txfonts}
\usepackage{longtable}
\usepackage{rotating}
\usepackage{natbib}
\usepackage{graphicx}
\usepackage{graphics}
\usepackage{psfrag}
\usepackage{amssymb}
\usepackage{xspace}
\usepackage{color}
\bibpunct{(}{)}{;}{a}{}{,}
\def\Teff{$T_{\mathrm{eff}}$}
\def\logg{\ensuremath{\log g}}

\def\vsini{\ensuremath{{\upsilon}\sin i}}
\def\kms{$\mathrm{km\,s}^{-1}$}

\def\loggf{log $gf$}

\def\ergscm{erg\,cm$^{-2}$\,s$^{-1}$}
\def\logR{\ensuremath{\log R^{\prime}_{\mathrm{HK}}}}
\def\Teq{$T_{\mathrm{eq}}$}
\def\So{$S_{\rm E=100}$}
\def\Smw{$S_{\rm MW}$}

\begin{document}
\title{The effect of ISM absorption on stellar activity measurements and its relevance for exoplanet studies}
\subtitle{}
\author{L. Fossati\inst{1,2} 			\and
	S. E. Marcelja\inst{1}			\and
	D. Staab\inst{2}			\and
	P. E. Cubillos\inst{1}			\and
	K. France\inst{3}			\and
	C. A. Haswell\inst{2}			\and
	S. Ingrassia\inst{4}			\and
	J. S. Jenkins\inst{5}			\and
	T. Koskinen\inst{6}			\and
	A. F. Lanza\inst{7}			\and
	S. Redfield\inst{8}			\and
	A. Youngblood\inst{3}			\and
	G. Pelzmann\inst{1,9}
}
\institute{
	Space Research Institute, Austrian Academy of Sciences, Schmiedlstrasse 		 6, A-8042 Graz, Austria\\
	\email{luca.fossati@oeaw.ac.at}
	\and
	School of Physical Sciences, The Open University, Walton Hall, Milton 			 Keynes MK7 6AA, UK
	\and
	Laboratory for Atmospheric and Space Physics, University of Colorado, 			 600 UCB, Boulder, CO 80309, USA	
	\and
	Department of Economics and Business - University of Catania, Corso 			 Italia, 55, I-95100 Catania, Italy
	\and
	Departamento de Astronomia, Universidad de Chile, Camino el Observatorio 	 1515, Las Condes, Santiago, Chile, Casilla 36-D
	\and
	Lunar and Planetary Laboratory, University of Arizona, 1629 East 			 University Boulevard, Tucson, AZ 85721-0092, USA
	\and
	INAF-Osservatorio Astrofisico di Catania, Via S.~Sofia, 78, I-95123 			 Catania, Italy
	\and
	Wesleyan University Astronomy Department, Van Vleck Observatory, 96 Foss 	 Hill Drive, Middletown, CT 06459, USA
	\and
	Institute for Physics/IGAM, NAWI Graz, Karl-Franzens-Universit\"at, 			 Universit\"atsplatz 5/II, 8010, Graz, Austria
}
\date{}
\abstract
{Past ultraviolet and optical observations of stars hosting close-in Jupiter-mass planets have shown that some of these stars present an anomalously low chromospheric activity, significantly below the basal level. For the hot Jupiter planet host WASP-13, observations have shown that the apparent lack of activity is possibly caused by absorption from the intervening interstellar medium (ISM). Inspired by this result, we study the effect of ISM absorption on activity measurements ($S$ and \logR\ indices) for main-sequence late-type stars. To this end, we employ synthetic stellar photospheric spectra combined with varying amounts of chromospheric emission and ISM absorption. We present the effect of ISM absorption on activity measurements by varying several instrumental (spectral resolution), stellar (projected rotational velocity, effective temperature, and chromospheric emission flux), and ISM parameters (relative velocity between stellar and ISM \ion{Ca}{ii} lines, broadening $b$-parameter, and \ion{Ca}{ii} column density). We find that for relative velocities between the stellar and ISM lines smaller than 30--40\,\kms\ and for ISM \ion{Ca}{ii} column densities $\log N_{\rm CaII}$\,$\gtrapprox$\,12, the ISM absorption has a significant influence on activity measurements. Direct measurements and three dimensional maps of the Galactic ISM absorption indicate that an ISM \ion{Ca}{ii} column density of $\log N_{\rm CaII}$\,=\,12 is typically reached by a distance of about 100\,pc along most sight lines. In particular, for a Sun-like star lying at a distance greater than 100\,pc, we expect a depression (bias) in the \logR\ value larger than 0.05--0.1\,dex, about the same size as the typical measurement and calibration uncertainties on this parameter. This work shows that the bias introduced by ISM absorption must always be considered when measuring activity for stars lying beyond 100\,pc. We also consider the effect of multiple ISM absorption components. We discuss the relevance of this result for exoplanet studies and revise the latest results on stellar activity versus planet surface gravity correlation. We finally describe methods with which it would be possible to account for ISM absorption in activity measurements and provide a code to roughly estimate the magnitude of the bias. Correcting for the ISM absorption bias may allow one to identify the origin of the anomaly in the activity measured for some planet-hosting stars.}
\titlerunning{The effect of ISM absorption on stellar activity measurements}
\authorrunning{L. Fossati et al.}
\maketitle
\section{Introduction}\label{sec:introduction}
Since the discovery of the first hot Jupiter (Jupiter-mass planets orbiting at less than 0.1\,AU from the host star), these planets were predicted to be affected by atmospheric escape \citep{mayor1995}. The detection of a 15\% dip in the Ly$\alpha$ transit light curve of HD\,209458b confirmed it \citep{vidal-madjar2003}. Escape processes, the upper atmosphere of planets, and star-planet interactions can be best studied observationally at ultraviolet (UV) wavelengths (at longer wavelengths the optical depth of the escaping material is low). So far, thorough UV observations have been published for three hot Jupiters \citep[HD\,209458b, HD\,189733b, and WASP-12b; e.g.][]{vidal-madjar2004,linsky2010,fossati2010a,lecavelier2012,haswell2012}, a warm-Neptune \citep[GJ\,436b; e.g.][]{kulow2014,ehrenreich2015}, and two super-Earths \citep[55\,Cnc\,e and HD\,97658b;][]{ehrenreich2012,bourrier2016}. The detection of the upper atmosphere of the hot Jupiter HD\,189733b has also been reported at optical wavelengths \citep{jensen2012,cauley2016}, though see \citet{barnes2016} for a cautionary note when interpreting transmission spectroscopy of lines which have a chromospheric emission component. These discoveries aided the development of sophisticated models of the upper atmosphere of close-in planets capable of treating various escape regimes \citep[e.g.][]{lammer2003,yelle2004,lecav2004,gm2007,mc2009,koskinen2014,salz2015,salz2016,erkaev2016,owen2016}.

The hot Jupiter WASP-12b orbits a late F-type star \citep{fossati2010b} and is one of the hottest (\Teq\,$\approx$\,2200\,K), most inflated known hot Jupiters, with an extremely short period \citep[about 1\,day;][]{hebb2009}. The near-UV transit observations, conducted with the Cosmic Origins Spectrograph (COS) on board the Hubble Space Telescope (HST), led to the detection of dips almost three times deeper than in the optical, revealing the presence of an escaping atmosphere \citep{fossati2010a,haswell2012}. The observations led to the detection in the planetary upper atmosphere of \ion{Mg}{ii} \citep{fossati2010a} and \ion{Fe}{ii} \citep{haswell2012}, the heaviest species yet detected in an exoplanet transit, and of a variable and early-ingress \citep{fossati2010a,haswell2012,nichols2015}, which stimulated a number of theoretical works \citep[e.g.][]{lai2010,vidotto2010,llama2011,bisikalo2013,alexander2016}.

The observations also revealed an anomaly in the stellar spectrum: a broad depression in place of the normally bright emission cores in the \ion{Mg}{ii}\,h\&k resonance lines \citep{haswell2012}. The anomaly is always present, regardless of the planet's orbital phase. \citet{haswell2012} analysed whether the anomalous line cores could be due to ({\it i}) intrinsically low stellar activity, or ({\it ii}) extrinsic absorption from either the interstellar medium (ISM) or from material within the WASP-12 system itself, presumably lost by the planet.

The comparison of optical spectra of WASP-12 with that of other distant and inactive stars revealed that WASP-12's \ion{Ca}{ii}\,H\&K line cores also show broad depressions similar to the \ion{Mg}{ii}\,h\&k line profiles \citep{fossati2013}. Measurements of the ISM absorption along the WASP-12 sight line showed that the ISM absorption is insufficient to produce these anomalies. Extrinsic absorption by material local to the WASP-12 system is therefore the most likely cause of the line core anomalies: gas escaping from the heavily irradiated planet \citep{haswell2012} and/or from evaporating rocky Trojan satellites \citep{kislyakova2016} could form a diffuse nebula enshrouding the entire system. Indirect observational and theoretical support to this scenario comes from the interpretation of the correlation between planetary surface gravity and stellar activity index \logR\ \citep{hartman2010,figueira2014,lanza2014,fossati2015a} and from 3D hydrodynamic simulations \citep[e.g.][]{carroll2016}.

For WASP-12, \citet{knutson2010} measured a \logR\ value of $-$5.5, which is well below the basal level of main-sequence solar-type stars of $-$5.1 \citep{wright2004}. This anomalously low \logR\ value is the direct consequence of the line core absorption, so similar activity index deficiencies might signal the presence of translucent circumstellar gas surrounding other stars hosting evaporating planets. Following this idea, \citet{fossati2013} and \citet{staab2016} identified about 20 other systems with an anomalously low activity index. Some of them are known hot Jupiter hosts, while others are candidates for dedicated follow-up studies to search for planets with escaping atmospheres.

The hot Jupiter planet host WASP-13 \citep{skillen2009} is one of the stars identified by \citet{fossati2013} as presenting an anomalously low \logR\ value. On the basis of COS far-UV spectra of WASP-13, \citet{fossati2015b} showed that, despite the anomalously low \logR\ value, the strength of the far-UV chromospheric emission lines is that of a typical middle-aged solar-type star, pointing toward the presence of significant extrinsic absorption. The further analysis of a high-resolution optical spectrum covering the \ion{Ca}{ii}\,H\&K resonance lines indicates that for WASP-13 the ISM absorption, as opposed to absorption in the immediate circumstellar environment, is the likely origin of the anomaly.

The vast majority of stars known to host a transiting planet lie rather far away from the solar system, in particular beyond 100--300\,pc, where the \ion{Ca}{ii} ISM column density is about 10$^{12}$\,cm$^{-2}$ or larger \citep[see Sect.~\ref{sec:discussion}; we note that for stars in the Kepler field, \ion{K}{i} ISM column densities have been derived by][]{johson2015}. For most of these stars, ISM absorption is likely to affect the shape of the emission core of the \ion{Ca}{ii}\,H\&K resonance lines, hence affecting the measurement of the \logR\ value. In particular, ISM absorption introduces a bias that leads to a \logR\ value that is systematically smaller compared to the unbiased one. We note that this bias does not affect the majority of the stars currently known to host planets detected by radial velocity due to their proximity to Earth.

In this work, we explore the effects of ISM absorption on measurements of the \ion{Ca}{ii}\,H\&K line core emission. Activity measurements, and the \logR\ value in particular, are currently used in the stellar and extra-solar planet (exoplanet) fields. Stellar activity measurements are of crucial importance for the study of late-type stars and, for distant objects, ISM absorption would lead to an underestimate of the activity and the amplitude of the activity cycle (see Sect.~\ref{sec:discussion}). In the exoplanet field, stellar activity measurements are used to constrain, for example, the amplitude of the intrinsic radial velocity jitter for planet searches \citep{haywood2016} and to study star-planet interactions \citep[e.g.][]{canto2011,miller2015}. It is therefore extremely important to be aware of the possible presence and magnitude of the bias introduced by ISM absorption in activity measurements.

This paper is organised as follows. Section~\ref{sec:input} describes the tools employed to carry out the analysis, while in Sect.~\ref{sec:resolution} we examine the effects of varying spectral resolution, normalisation, and background subtraction on activity measurements. In Sects.~\ref{sec:ISM_general_sun} and \ref{sec:ISM_general_all} we explore the effect of ISM absorption on activity measurements first for a Sun-like star and then extend it to other stellar types. In Sect.~\ref{sec:discussion} we discuss the results and present the conclusions in Sect.~\ref{sec:conclusion}.
\section{Input parameters and activity measurements}\label{sec:input}
This work is based on synthetic photospheric stellar spectra calculated in local thermodynamical equilibrium (LTE) employing the LLmodels stellar atmosphere code \citep{llm}. We computed models for the Sun \citep[i.e. effective temperature \Teff\ of 5777\,K, surface gravity \logg\ of 4.438, microturbulence velocity of 0.875\,\kms, and abundances from][]{asplund2009} and for stars with \Teff\ ranging between 4000 and 6500\,K, in steps of 100\,K, \logg\ fixed at a value of 4.4, to simulate main-sequence late-type stars, a microturbulence velocity of 1.0\,\kms, and solar abundances. The synthetic fluxes were calculated with a sampling of 0.005\,\AA\ bin width in the region covered by the \ion{Ca}{ii}\,H\&K lines. The full-width-half-maximum of the chromospheric emission lines is about 0.15\,\AA, while that of the photospheric \ion{Ca}{ii}\,H\&K lines is several Angstroms. In Sects.~\ref{sec:resolution} and \ref{sec:ISM_general_sun}, we present in detail the results obtained when considering the synthetic spectrum of the Sun, while in Sect.~\ref{sec:ISM_general_all} we extend the main results to the other simulated stars. Throughout, we do not include any rotational and macroturbulent broadening in the stellar spectra unless explicitly stated.

Since LLmodels computes only photospheric fluxes, we artificially added the chromospheric emission in the core of the \ion{Ca}{ii}\,H\&K lines using Gaussian profiles broadened to match the line width of the solar emission line profiles \citep[e.g.][]{fossati2015b}. We control the line core emission strength through a parameter $E$ (in units of \ergscm), which is the disk-integrated \ion{Ca}{ii} emission flux at 1\,AU and we distribute the emission between the H and K lines according to their relative strengths (i.e. following their \loggf\ values). The parameter $E$ is the same provided by the empirical relations of \citet{linsky2013}. For reference, the average solar $S$-index is 0.176 \citep{mamajek2008} corresponding to an $E$ value of 25.53\,\ergscm, which is very close to the value of 24.67$\pm$0.42\,\ergscm\ obtained by \citet{oranje1983} from spectroscopic observations of the Sun as a star between 1979 and 1981. At solar maximum and minimum the $S$ values are respectively 0.226 and 0.146, which correspond to $E$ values of 41.9 and 15.7\,\ergscm.

The \logR\ parameter gives a measure of the emission present in the core of the \ion{Ca}{ii}\,H\&K lines, independent of the stellar effective temperature. The \logR\ value is derived from a calibration of the S-index ($S$) \citep{noyes1984,mittag2013} that is calculated as
\begin{equation}
S=\alpha\frac{H+K}{R+V}\,,
\label{eq:s}
\end{equation}
where $\alpha$ is an historical (dimensionless) conversion factor \citep{hall2007}, $H$ and $K$ are fluxes within triangular bandpasses covering the \ion{Ca}{ii}\,H\&K line cores, and $R$ and $V$ are fluxes within continuum windows on either side of the \ion{Ca}{ii} lines, as shown for example in Fig.~1 of \citet{jenkins2006}. The centres and widths (in \AA) of the $H$, $K$, $R$, and $V$ bands are 3968.470, 3933.664, 4001.07, 3901.07\,\AA\ and 1.09, 1.09, 10.0, 10.0\,\AA, respectively. To compute the $S$ index, we calculate the mean flux in each bandpass instead of using integrated fluxes. This reduces edge effects from finite sampling at the boundaries of the four bands and it leads to $\alpha$\,$\approx$\,1 \citep{lovis2011}. The fact that the real line core emission observed for the Sun is double peaked, while we simulate the emission with a single Gaussian profile, has a small effect on the measurement of the $S$ values because the $H$ and $K$ bandpasses are broader than the emission features. We tested this by looking at the difference in the $S$ value for a Sun-line star where the line core emission was simulated with a single- or double-peaked Gaussian, obtaining $S_{\rm single}-S_{\rm double}$\,=\,0.0015. We remark that the true chromospheric emission lines are best reproduced by partial frequency redistribution non-LTE line profiles that lead to small departures from the assumed LTE photospheric profile plus Gaussian emission and should not significantly affect the results of this work. This is confirmed by the agreement we obtain for the solar chromospheric flux measured from the $S$ index and that given by \citet{oranje1983}. We, however, come back to this point in Sect.~\ref{sec:varRV_fixN}.

\begin{figure}[]
\includegraphics[width=90mm,clip]{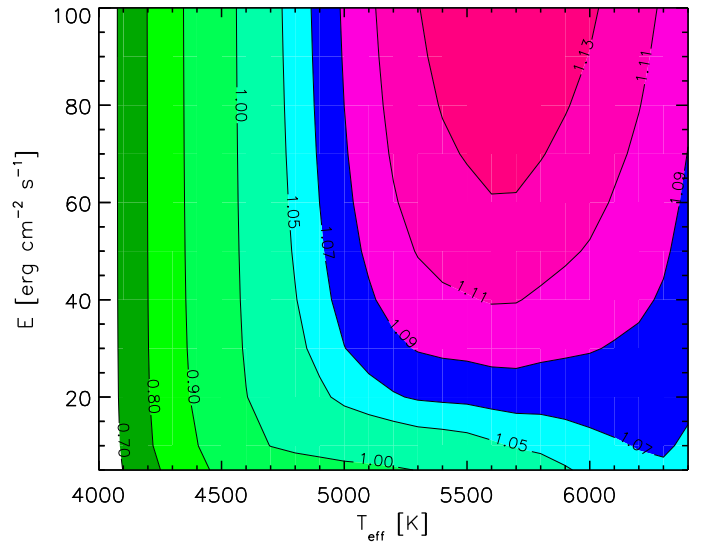}
\caption{Conversion factor $\beta$ between $S$ and \Smw\ as a function of \Teff\ and $E$. The values along the line of each contour quantify $\beta$.}
\label{fig:beta}
\end{figure}

By following the method described by \citet{middelkoop1982} and \citet{rutten1984}, or in Sect.~3.3 of \citet{mittag2013}, we concluded that our calculated $S$ index is not in the Mount Wilson system (\Smw). We therefore derived the conversion factor $\beta$ between $S$ and \Smw\ using the relation \citep[Eq.~2 of][]{mittag2013}
\begin{equation}
S\,\beta = S_{\rm MW} = \alpha\frac{\mathcal{F}_{\rm HK}}{\mathcal{F}_{\rm RV}} = \alpha\frac{E'+\mathcal{F}_{\rm HK,phot}}{\mathcal{F}_{\rm RV}}\,,
\label{eq:sMW}
\end{equation}
where $\mathcal{F}_{\rm HK}$ is the total stellar surface flux (i.e. photospheric plus chromospheric not weighted by the triangular bandpasses) in the core of the H and K lines, $\mathcal{F}_{\rm RV}$ is the stellar photospheric flux in the $R$ and $V$ continuum bands, $E'$ is the chromospheric stellar surface flux (i.e. $E'=E\times1{\rm AU}^2/R_{\rm star}^2$), and $\mathcal{F}_{\rm HK,phot}$ is the photospheric stellar surface flux in the core of the H and K lines. Following \citet{mittag2013}, $\mathcal{F}_{\rm RV}$ and $\mathcal{F}_{\rm HK,phot}$ are a function of the $B-V$ colour as follows
\begin{equation}
\log{\frac{\mathcal{F}_{\rm RV}}{\alpha}} = 8.25-1.67(B-V)
\label{eq:Frv}
\end{equation}
and
\begin{equation}
\log{\mathcal{F}_{\rm HK,phot}} = 7.49-2.06(B-V)\,.
\label{eq:Fhkphot}
\end{equation}
The expressions given in Eqs.~\ref{eq:Frv} and \ref{eq:Fhkphot} are valid for main-sequence stars with 0.44\,$\leq$\,$B-V$\,$\leq$\,1.28 (i.e. 4000\,$\lesssim$\,\Teff\,$\lesssim$\,6300\,K). For cooler stars, $B-V$ is no longer a good measure of \Teff. Since $\beta$ depends on $B-V$ (hence \Teff) and $E$, we mapped the value of $\beta$ in the parameter space considered in this work (Fig.~\ref{fig:beta}) and used it to convert our results into the Mount Wilson system. The $B-V$ values have been derived from the \Teff\ values interpolating Table~B1 of \citet{gray2005}.

Since the parameter $E$ is not of common use for activity measurements, we give in Eq.~\ref{eq:lin.relation} the linear relationship between $E$ and \Smw\
\begin{equation}
S_{\rm MW} = 0.097925 + E \times 0.003058\,.
\label{eq:lin.relation}
\end{equation}
For several of the tests presented below, we use the \Smw\ value corresponding to a chromospheric emission of $E$\,=\,100\,\ergscm\ and no ISM absorption. Following Eq.~\ref{eq:lin.relation}, this emission value corresponds to \Smw\,=\,0.404.

\begin{figure*}[]
\includegraphics[width=185mm,clip]{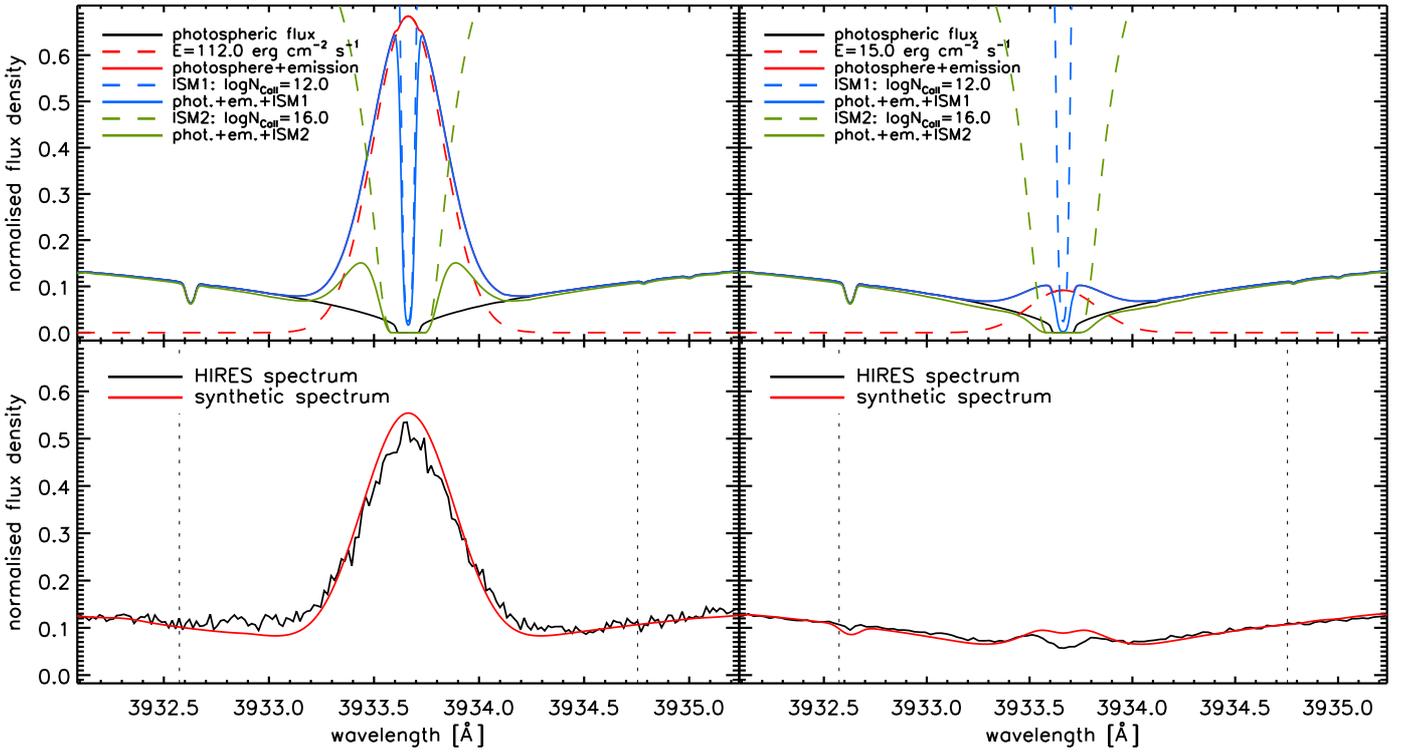}
\caption{Top-left: LLmodels photospheric spectrum of the Sun in the region of the \ion{Ca}{ii}\,K line core (black solid line) modified by adding a line core emission of 112\,\ergscm\ (red solid and dashed lines). This composite spectrum has then been modified by two ISM absorption features, located at the line centre, calculated with \ion{Ca}{ii} column densities of 10$^{12}$ and 10$^{16}$\,cm$^{-2}$ (blue and green dashed and solid lines). Bottom-left: comparison between a high resolution HIRES spectrum of the young Sun-like star EK\,Dra and the corresponding synthetic spectrum obtained by adding a line core emission of $E$\,=\,112\,\ergscm\ to the photospheric spectrum. The two spectra return the same \Smw\ value of 0.440. The synthetic spectrum accounts also for the instrumental spectral resolution and the stellar projected rotational velocity ($\approx$15\,\kms). Right: same as the left panels, but for the hyper-velocity halo Sun-like star HD\,29587. In this case, the emission flux is 15\,\ergscm, corresponding to a \Smw\ value of 0.144. The synthetic spectrum does not match the line profile in detail, possibly due to non-LTE effects, but it reproduces the empirical $S$ value. In the top panels, the red solid lines are visible just in the very core of the line, as it is otherwise covered by the blue solid line.}
\label{fig:synthetic}
\end{figure*}

To illustrate the interplay between the photospheric and chromospheric fluxes and the ISM absorption component, we present in Figure~\ref{fig:synthetic} two examples of how the line core emission and ISM absorption modify the initial photospheric spectrum. We also show here comparisons to the observed spectra of EK\,Dra (a very active nearby Sun-like star) and HD\,29587 (a hyper-velocity halo Sun-like star), obtained with HIRES at KECK observatory, and the corresponding synthetic spectra obtained by adding to the photospheric spectrum a line core emission of $E$\,$\approx$\,112 and 15\,\ergscm, respectively. We chose these stars because they lack ISM absorption lines close to the \ion{Ca}{ii}\,H\&K line cores. The $E$ values are those that lead to the same \Smw\ values measured from the two spectra.

\begin{figure}[]
\includegraphics[width=90mm,clip]{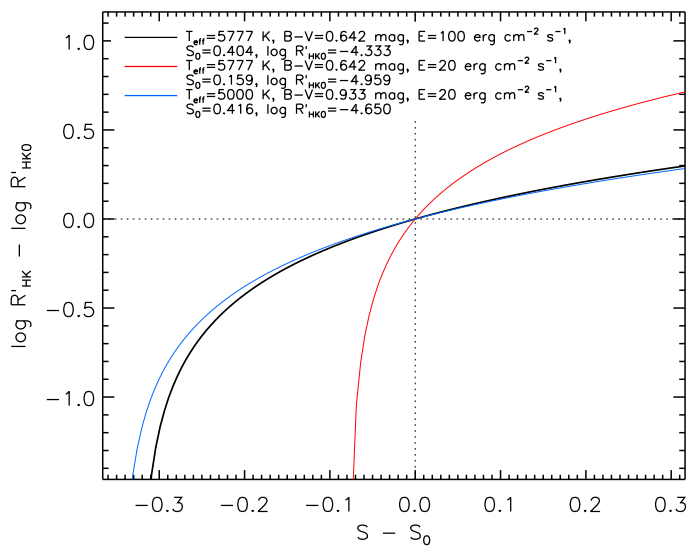}
\caption{Relation between differences in the \Smw\ and \logR\ values for a Sun-like star with a chromospheric emission of 40 (red) and 100\,\ergscm\ (black) and a star with \Teff\,=\,5000\,K and a chromospheric emission of 40\,\ergscm\ (blue). The legend indicates the parameters used for the computations. S$_{\rm 0}$ and $\log R^{\prime}_{\mathrm{HK0}}$ are respectively the \Smw\ and \logR\ values obtained for the emission given in the legend.}
\label{fig:conversion}
\end{figure}

Despite the possibility to directly convert the \Smw\ values into \logR\ values, in this work we consider mostly the \Smw\ values, which are the actual measured quantities, while the \logR\ values are derived through calibrations. Nevertheless, to give an idea of how a variation in the \Smw\ value maps into \logR, we show in Fig.~\ref{fig:conversion} the relation between differences in the \Smw\ and \logR\ values for three different cases: a Sun-like star with a chromospheric emission of 20 and 100\,\ergscm, and a star with \Teff\,=\,5000\,K with a chromospheric emission of 20\,\ergscm. We chose these three cases to highlight the effects of changing $E$ and stellar spectral type.

For completeness, we describe here how we converted the \Smw\ values into \logR. For this operation, we followed the procedure described by \citet{noyes1984} and converted the \Teff\ values into $B-V$ colours interpolating from Table~B1 of \citet{gray2005}. The range of validity of this conversion is the same as that of Eqs.~\ref{eq:Frv} and \ref{eq:Fhkphot}. For the Sun, we adopted a $B-V$ colour of 0.642\,mag \citep{cayrel1996}. In practice, we converted the \Smw\ values into \logR\ as follows. We first derived the $R_{\rm HK}$ parameter as \citep{noyes1984}
\begin{equation}
R_{\rm HK} = 1.34\times10^{-4} \times CF \times S_{\rm MW}\,,
\label{eq:RHK}
\end{equation}
where, for main-sequence stars, $CF$ is \citep{rutten1984}
\begin{equation}
\log{CF} = 0.25\times(B-V)^3 - 1.33\times(B-V)^2 + 0.43\times(B-V) + 0.24\,.
\label{eq:CF}
\end{equation}
Following \citet{noyes1984}, for main-sequence stars the photospheric flux contribution in the core of the \ion{Ca}{ii} lines is
\begin{equation}
\log{R_{\rm HK,phot}} = -4.02 - 1.40\times(B-V)\,.
\label{eq:RHKphot1}
\end{equation}
We note that this expression has been revised by \citet{mittag2013} using the PHOENIX stellar atmosphere code as
\begin{equation}
R_{\rm HK,phot} = \frac{\mathcal{F}_{\rm HK,phot}}{\sigma T_{\rm eff}^4}\,,
\label{eq:RHKphot2}
\end{equation}
where for main-sequence stars $\mathcal{F}_{\rm HK,phot}$ is that given in Eq.~\ref{eq:Fhkphot}. Finally, the \logR\ value is obtained from
\begin{equation}
\log{R^{\prime}_{\mathrm{HK}}} = \log{(R_{\rm HK} - R_{\rm HK,phot})}\,.
\label{eq:R'HK}
\end{equation}
Figure~7 of \citet{staab2016} shows the observed distribution of \logR\ values, as a function of effective temperature, for main-sequence stars.
%
\section{Effects of spectral resolution, background subtraction, and normalisation on the $S$ index of a Sun-like star}\label{sec:resolution}
As shown in Sect.~\ref{sec:discussion}, the stars for which the effect of ISM absorption on activity measurements is relevant lie relatively far away from Earth (i.e. beyond 100\,pc). Because of the large distances, these stars are typically faint and it is difficult to obtain high quality spectra, particularly in the region of the \ion{Ca}{ii} lines. For this reason, the spectral background subtraction may not always be optimal. Therefore, we analyse here the effects of an under- or over-subtraction of the background on activity measurements. We also take advantage of the freedom given by synthetic spectra to analyse the effects on the \Smw\ value of varying spectral resolution and normalisation.

\begin{figure}[]
\includegraphics[width=90mm,clip]{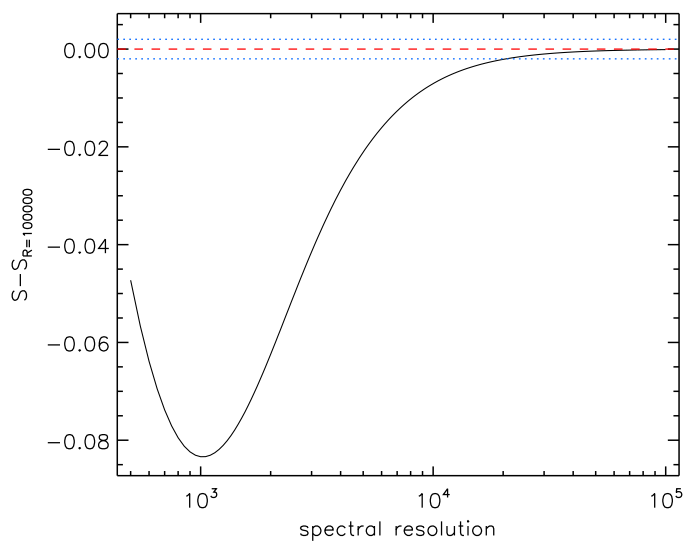}
\caption{Deviation of the \Smw\ value from that at R\,=\,100\,0000 as a function of spectral resolution. The dotted lines indicate typical photon noise uncertainties on \Smw\ (i.e. 0.002).}
\label{fig:Sagainst_resolution}
\end{figure}

To study the impact of spectral resolution, we added an emission $E$\,=\,100\,\ergscm\ to the synthetic photospheric spectrum of the Sun and convolved it with Gaussians to reach resolving powers ranging from 500 to 100\,000 in steps of 100. Figure~\ref{fig:Sagainst_resolution} shows how \Smw\ varies as a function of spectral resolution and compares it to typical photon-noise uncertainties obtained from measurements of \Smw\ conducted on real spectra \citep[signal-to-noise ratio of the spectra of about 50; e.g.][]{jenkins2008}. The measurement of the $S$-index is not affected by the instrumental resolving power down to values of about 20\,000. In the case of real stars and observed spectra, the spectral resolution has actually a small effect on activity measurements. This is because the calibration onto the Mount Wilson system, carried out using calibration stars, corrects for most of the distortions possibly caused by a low resolution \citep{jenkins2011}.

\begin{figure}[]
\includegraphics[width=90mm,clip]{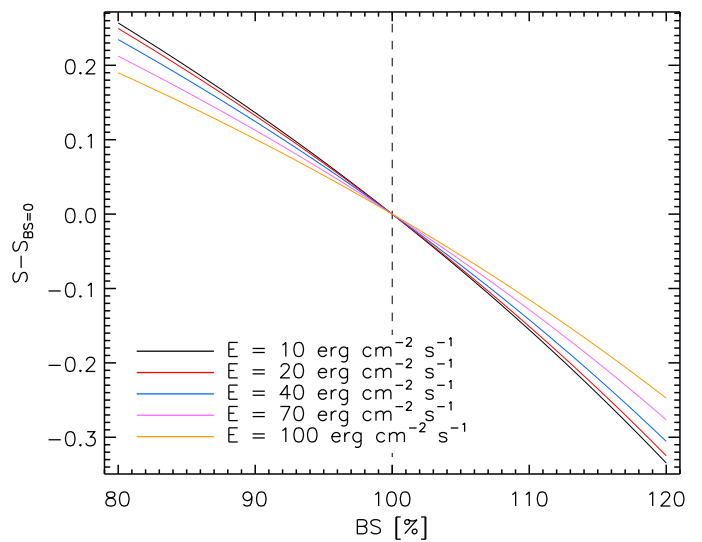}
\caption{Deviation of the $S$ value from that derived with perfect background subtraction as a function of the deviation from a perfect background subtraction, considering five different chromospheric emission levels.}
\label{fig:Sagainst_scatter}
\end{figure}

A more important effect for measurements of the $S$ index from real stellar spectra is that of a non-optimal removal of the background during spectral reduction (e.g. residual instrumental scattered light, non-optimal removal of the detector dark level). This is particularly important in the case of low signal-to-noise spectra for which the background subtraction is difficult and has a large impact on the extracted spectrum. Figure~\ref{fig:Sagainst_scatter} shows the effect of a non-optimal background subtraction on the measurement of the $S$ value, assuming five different chromospheric emission levels. We implemented non-optimal background subtraction in the synthetic spectra using the following formula
\begin{equation}
\label{eq:BS}
Syn_{BS}=1-[BS\times(1-Syn)]/100\,\,,
\end{equation}
where $Syn_{BS}$ is the modified synthetic spectrum, $BS$ is the amount of non-optimal background subtraction (in \%), and $Syn$ is the initial (i.e. unmodified) synthetic spectrum. As expected, an over-subtraction of the background ($BS$ values greater than 100\%) decreases the flux in the \ion{Ca}{ii}\,H\&K line cores, compared to that of the $R$ and $V$ bands, leading to a decrease of the $S$ value. The opposite occurs with an under-subtraction of the background, or equivalently the presence of residual scattered light. Figure~\ref{fig:Sagainst_scatter} shows that the impact of background subtraction on the $S$ value depends also on the emission level.

We also checked the sensitivity of the $S$ index to the normalisation procedure, hence considering un-normalised spectra, as is the case for most measurements in the literature. We multiplied or divided a spectrum to which we added a chromospheric emission of $E$\,=\,100\,\ergscm\ by polynomials of 1st, 2nd, and 3rd degree, also varying their coefficients, always obtaining an $S$ value within 0.002 of the reference. This agrees with, for example, \citet{gray2003}.
\section{Effects of ISM absorption on the $S$ index: the case of a Sun-like star}\label{sec:ISM_general_sun}
We present here the effects of ISM absorption on \Smw, varying both the ISM radial velocity (RV)\footnote{We consider the stellar spectrum to be in the laboratory reference frame.} and the \ion{Ca}{ii} column density ($\log N_{\rm CaII}$), and considering the synthetic fluxes calculated for a Sun-like star. We modelled the ISM absorption using Voigt profiles whose strength is controlled by the \ion{Ca}{ii} column density and a broadening $b$-parameter of 2.0\,\kms. We adopted this value following the results of \citet{redfield2002}, \citet{redfield2004}, and \citet{wyman2013} who measured \ion{Ca}{ii} ISM absorption lines for stars within 100\,pc obtaining average $b$-parameters below 2.5\,\kms. As a further confirmation, \citet{welty1996}, for stars lying between 100\,pc and 1\,kpc, measured broadening $b$-parameters ranging predominantly between 0.5 and 2.5\,\kms. However, in Sect.~\ref{sec:varN_varRV} we explore the effects of varying the $b$-parameter. Throughout, we consider the effect of just one ISM absorption component, but discuss, however, the case of multiple ISM components in Sect.~\ref{sec:multiple}.
\subsection{Variable ISM radial velocity and fixed column density}\label{sec:varRV_fixN}
We first looked at the effect of varying the radial velocity of the ISM absorption line keeping $E$ and $\log N_{\rm CaII}$ fixed at 100\,\ergscm\ and 12, respectively. We varied RV from $-$120.0 to $+$120\,\kms, in steps of 0.5\,\kms, recalculating \Smw\ at each position of the ISM absorption line. When compared to the case of observed spectra, the RV is the difference between the stellar and ISM radial velocities.

\begin{figure}[]
\includegraphics[width=90mm,clip]{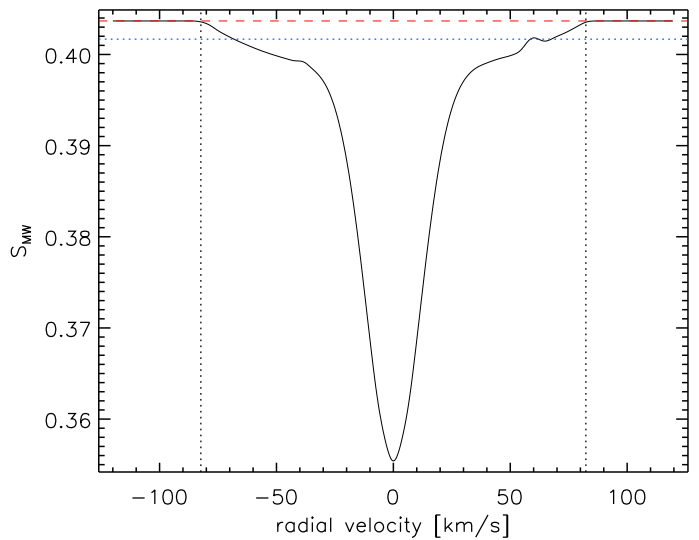}
\caption{Variation of \Smw\ as a function of radial velocity of the ISM absorption line. The dashed red line shows the \So\ value and the horizontal dotted blue line indicates the typical photon noise uncertainty, while the vertical dotted black lines represent the boundaries of the $H$ and $K$ passbands used to calculate the $S$ value, converted to velocity space.}
\label{fig:changing_vr}
\end{figure}

Figure~\ref{fig:changing_vr} shows how \Smw\ varies as a function of the ISM radial velocity. The general shape of the variation drawn in Fig.~\ref{fig:changing_vr} is driven by the fact that the effect of the absorption on \Smw\ depends on the relative strengths between the stellar flux (photospheric plus chromospheric) and the ISM absorption itself. In particular, the ISM has a larger impact on \Smw\ when the absorption line is located in a region where the stellar fluxes are higher and where the triangular core bandpasses peak.

At large velocities, the ISM absorption line lies outside of the band used to measure the $S$ index, which is why there is no variation beyond $\pm$86\,\kms. Within these limits, the ISM absorption significantly reduces the $S$ value with the maximum obtained at RV\,=\,0\,\kms. However, the curve shown in Fig.~\ref{fig:changing_vr} should in reality be slightly asymmetric with a double-peaked minimum because of the real shape of the chromospheric emission (see for example the bottom-right panel of Fig.~\ref{fig:synthetic}), while we approximate the chromospheric emission with a single Gaussian (this consideration holds also for the contour plots shown in the next Sections). Deviations from the Gaussian profile are, however, expected to be small. The bump located at a radial velocity of about $+$60\,\kms\ is caused by the presence of a strong photospheric absorption line just redwards of the core of the \ion{Ca}{ii}\,H line.
%
\subsection{Variable ISM column density and fixed radial velocity}\label{sec:varN_fixRV}
\begin{figure}[]
\includegraphics[width=90mm,clip]{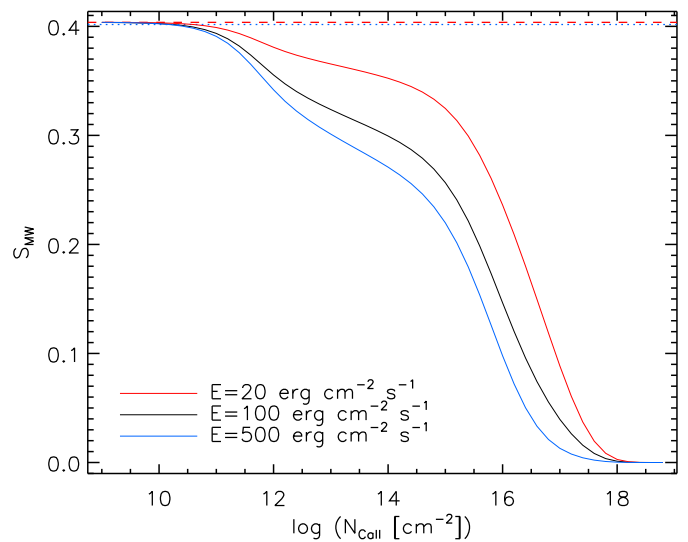}
\caption{Variation of \Smw\ as a function of $\log N_{\rm CaII}$. Each line has been obtained considering different values of the emission $E$, as indicated in the legend. The dashed line shows the value of \So, while the dotted line indicates typical photon noise uncertainties. For comparison purposes, the blue and red lines have been rescaled in order to match the black line at $\log N_{\rm CaII}$\,=\,0.}
\label{fig:changing_N}
\end{figure}

To explore the effect of the amount of ISM absorption on \Smw, we fixed $E$ equal to 20, 100, or 500\,\ergscm\ and the radial velocity of the ISM absorption line equal to 0.0\,\kms, varying the ISM column density $\log N_{\rm CaII}$ from 9.0 to 19.0\footnote{This is the value at which a line core emission of 100\,\ergscm\ is completely absorbed by the ISM.}, in steps of 0.2, recalculating \Smw\ at each step. The result is shown in Fig.~\ref{fig:changing_N}. As expected, the value of \Smw\ decreases with increasing column density. Already at $\log N_{\rm CaII}$\,$\approx$\,10.5 the deviation becomes significant compared to typical photon noise uncertainties. The bias on \Smw\ caused by ISM absorption increases with increasing chromospheric emission.

With increasing ISM column density, \Smw\ decreases because the ISM line absorbs the chromospheric emission. The bend present at $\log N_{\rm CaII}$\,$\approx$\,14.0--15.0 is caused by the saturation of both (H and K) ISM absorption lines. For larger values of $\log N_{\rm CaII}$, \Smw\ decreases monotonically with increasing $\log N_{\rm CaII}$ until the emission (photospheric and chromospheric) in both line cores is completely absorbed. Figure~\ref{fig:changing_N} also shows that the effect of the ISM absorption on \Smw\ increases with increasing line core emission.
%
\subsection{Saturation}
\begin{figure}[]
\includegraphics[width=90mm,clip]{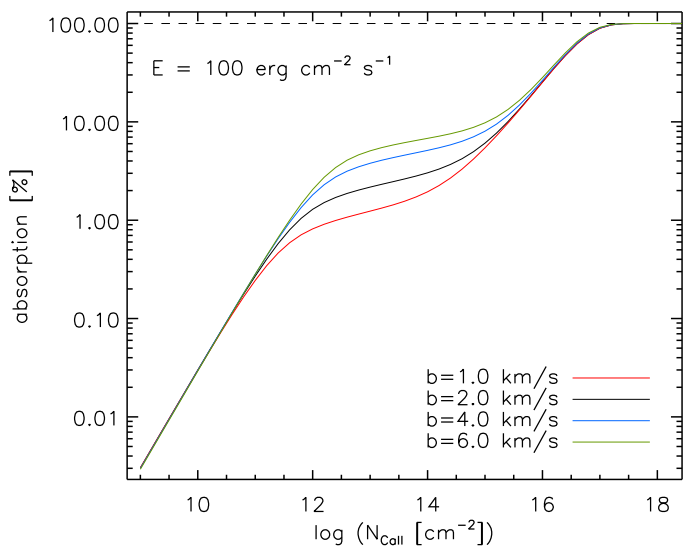}
\caption{Absorption (in \%) caused by one ISM line as a function of \ion{Ca}{ii} column density assuming four different values of the broadening $b$-parameter. The chromospheric emission is $E$\,=\,100\,\ergscm. The y-axis is in log scale.}
\label{fig:saturation}
\end{figure}

Line saturation plays an important role in shaping the effect of ISM absorption on activity measurements. Figure~\ref{fig:saturation} shows the absorption (in \%) caused by various amounts of \ion{Ca}{ii} ISM column density (one absorption line) on a chromospheric emission of $E$\,=\,100\,\ergscm, assuming four different values of the broadening $b$-parameter. The absorption increases linearly up to a column density of $\log N_{\rm CaII}$\,$\approx$\,11.5--12, where the absorption saturates and depends strongly on the broadening $b$-parameter, up to $\log N_{\rm CaII}$\,$\approx$\,15--16. With increasing column density, the absorption increases as the square root of $\log N_{\rm CaII}$. At $\log N_{\rm CaII}$\,$\approx$\,17, the chromospheric emission is completely absorbed.
%
\subsection{Variable ISM column density and radial velocity}\label{sec:varN_varRV}
In the remainder of the paper, we consider ISM \ion{Ca}{ii} column densities ranging between 10$^{9}$ and 10$^{14}$\,cm$^{-2}$, which is roughly the range expected for stars closer than 500-600\,pc (see Sect.~\ref{sec:discussion}). Figure~\ref{fig:changing_N_VR_R} shows the differences between the reference $S$ value \So\  (i.e. without ISM absorption) and the \Smw\ values obtained by changing the spectral resolution (100\,000, 65\,000, 45\,000, 20\,000, 10\,000, and 5\,000) and the ISM parameters (RV and $\log N_{\rm CaII}$), keeping $E$ fixed at 100\,\ergscm. This plot indicates that the spectral resolution does not dramatically affect the results, allowing us to fix the spectral resolution to 100\,000 and hence to concentrate on the other parameters. However, unless the chromospheric emission is particularly strong and/or the ISM column density is particularly high, below a resolution of about 10\,000, it becomes difficult to visually identify the presence of an ISM absorption line, even with a spectrum of very high quality.

In Fig.~\ref{fig:changing_N_VR_vsini}, we explore the effect of varying stellar projected rotational velocity (\vsini) and determine that it has a small impact on the differences between \So\ and \Smw\ compared to other effects. We therefore maintain the assumption of no rotation.

\begin{figure*}[ht!]
\includegraphics[width=185mm,clip]{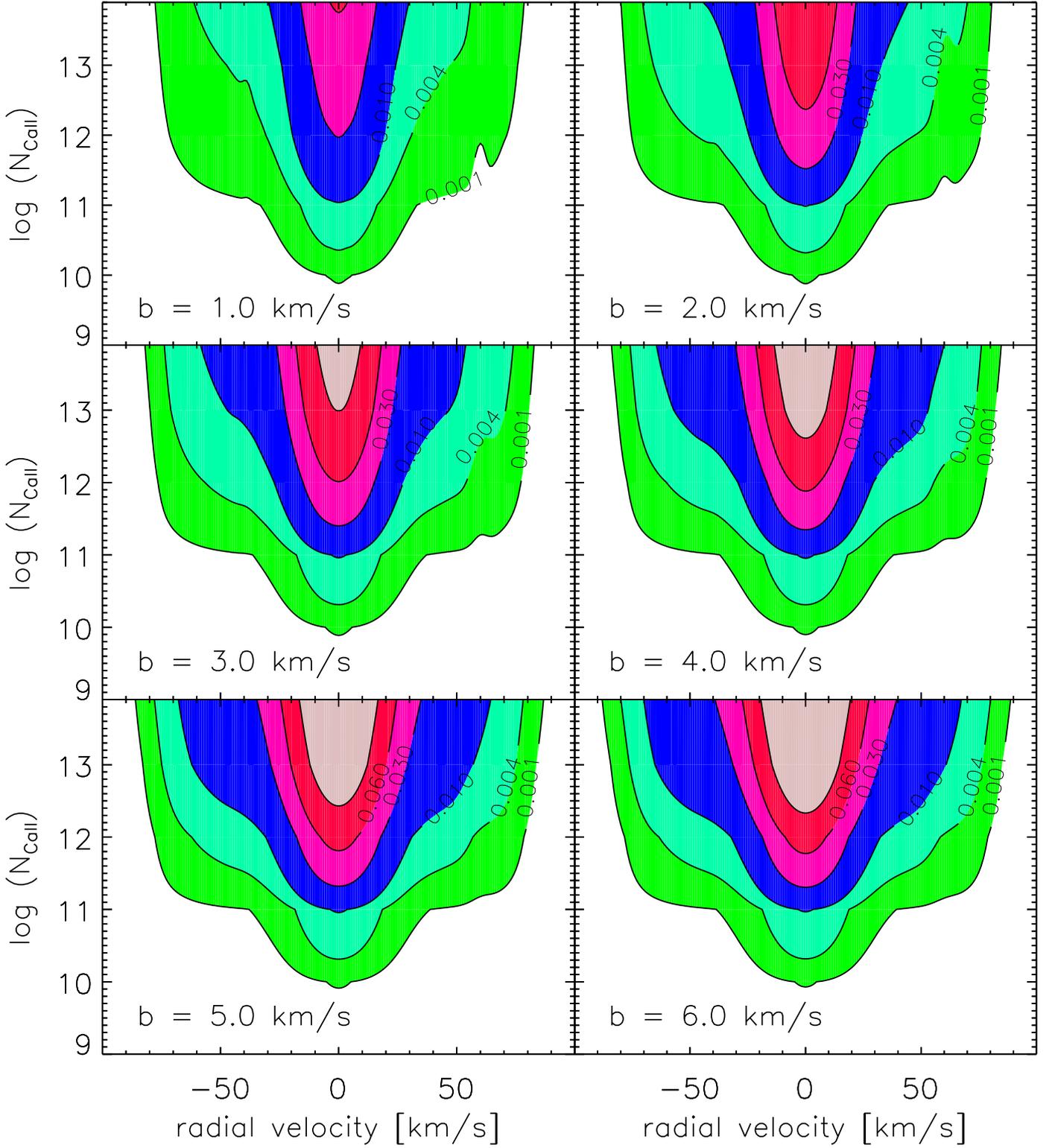}
\caption{Difference between \So\ and \Smw\ obtained by varying the ISM broadening ($b$-parameter), radial velocity (RV), and column density ($\log N_{\rm CaII}$), keeping $E$ fixed at 100\,\ergscm. The values along the lines of each contour (0.001, 0.004, 0.01, 0.03, 0.06, 0.11) quantify the \So\,$-$\,\Smw\ difference.}
\label{fig:changing_N_VR_b}
\end{figure*}

In Fig.~\ref{fig:changing_N_VR_b}, we show instead the effect of varying the ISM broadening ($b$-parameter), radial velocity, and column density, keeping $E$ fixed at 100\,\ergscm. As one may expect, by increasing the broadening of the ISM absorption line, the bias on \Smw\ increases as well. In particular, by doubling the $b$-parameter, the systematic decrease of \Smw\ increases by about a factor of two.

\begin{figure*}[ht!]
\includegraphics[width=90mm,clip]{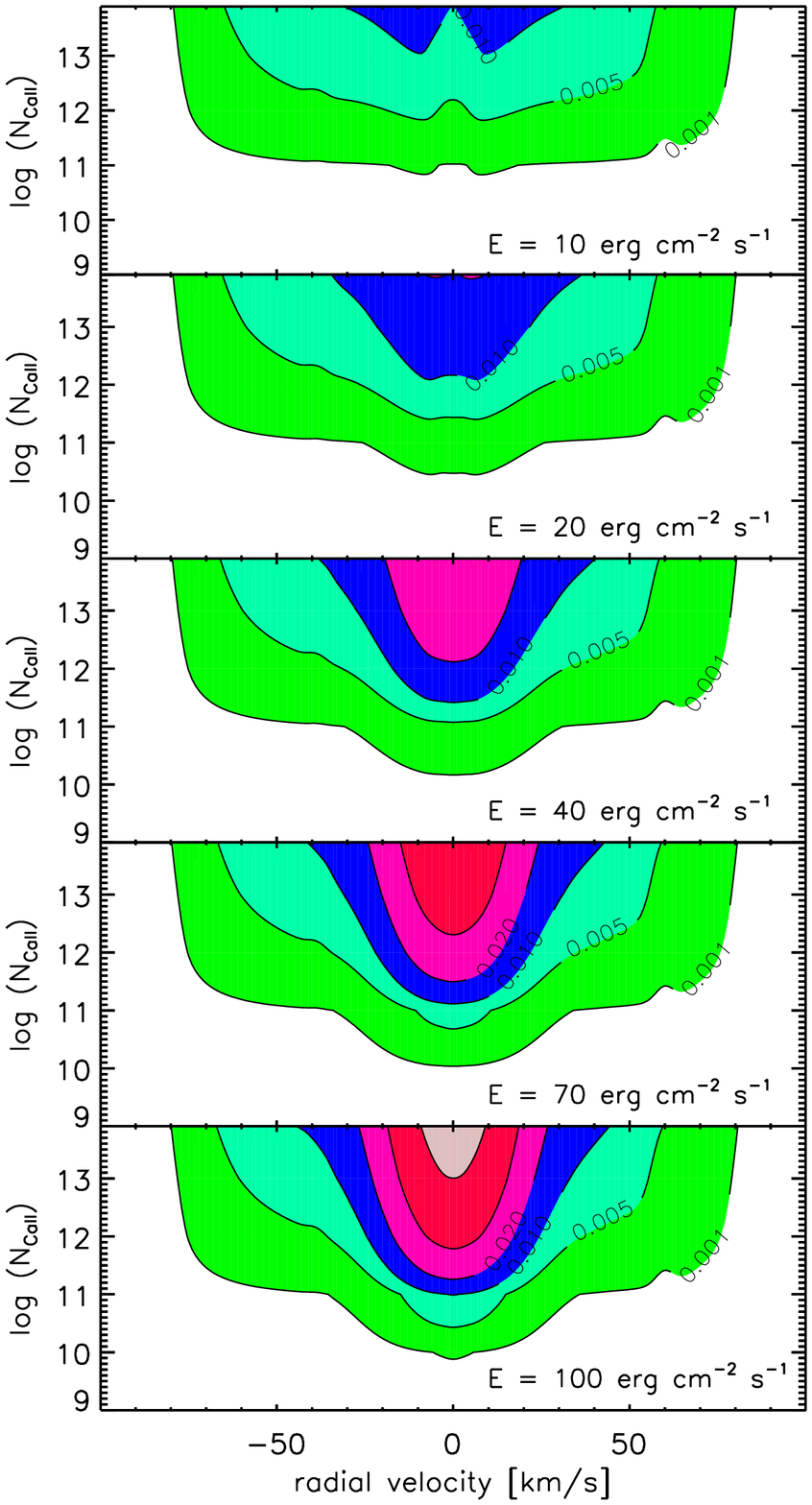}
\hspace{1cm}
\includegraphics[width=90mm,clip]{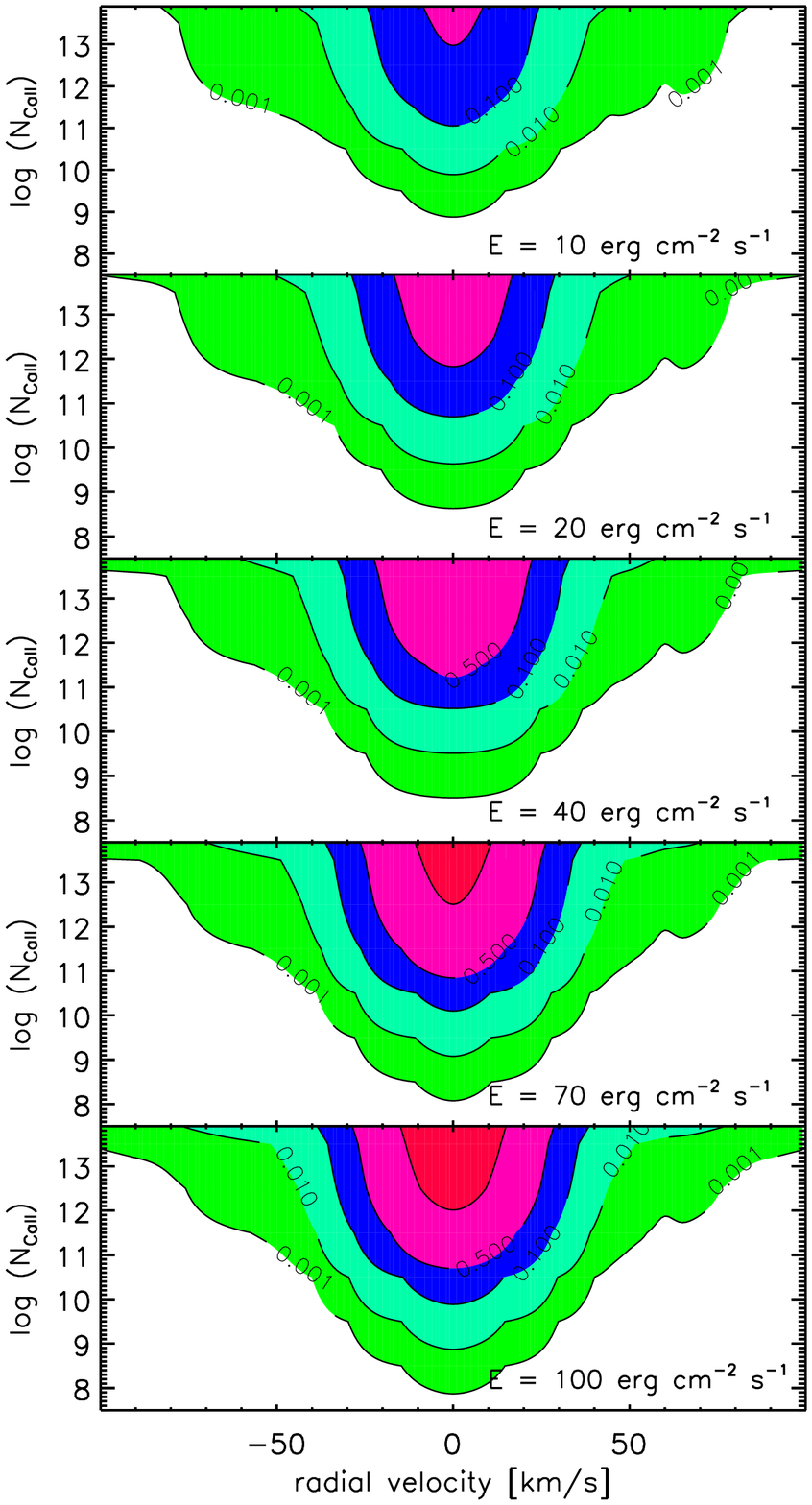}
\caption{Left: differences between the reference $S$ value \So\ calculated without ISM absorption, and \Smw\ obtained by varying the ISM parameters for five different values of $E$. The adopted spectral resolution is 100\,000. The values along the line of each contour (0.001, 0.005, 0.01, 0.02, 0.04, 0.08) quantify the \So\,$-$\,\Smw\ difference. The results shown in the third and fifth panels can be directly converted into differences in \logR\ values using respectively the red and black lines displayed in Fig.~\ref{fig:conversion}. Right: same as the left panel, but for a star with \Teff\,=\,4000\,K. The values along the line of each contour are: 0.001, 0.01, 0.1, 0.5, 3.0.}
\label{fig:changing_N_VR_Ca2em}
\end{figure*}

The left panel of Fig.~\ref{fig:changing_N_VR_Ca2em} shows the differences between \So, calculated for different values of the chromospheric emission $E$, and \Smw\ obtained by varying the ISM radial velocity and column density (the $b$-parameter is fixed at 2\,\kms). The left panel of Fig.~\ref{fig:changing_N_VR_Ca2em} shows that the effect of the ISM absorption on \Smw\ increases with increasing line core emission. This is most evident at RV close to 0\,\kms\ and at large $\log N_{\rm CaII}$ values, where the differences between the five panels are caused by the fact that a smaller ISM absorption is necessary to absorb a weaker emission.
%
\subsection{Multiple ISM components}\label{sec:multiple}
We tested the effect of the presence of two (or more) ISM absorption lines on \Smw. We did this by considering one absorption line calculated for a value of the \ion{Ca}{ii} column density that does not saturate the line (i.e. $\log N_{\rm CaII}$\,$<$\,12.0) and by adding further absorption lines with $\log N_{\rm CaII}$ varying between 9 and 19. If the lines are at the same radial velocity and are not saturated, their contribution is additive and there is no difference from the case of a single line. Differences are present if the absorption lines do not have the same radial velocity. In this case, the effect of the ISM absorption on \Smw\ depends not only on the strength of the absorption lines, but also on their position with respect to the core of the \ion{Ca}{ii}\,H\&K lines. If one of the ISM absorption lines is saturated, the contribution of the two lines is not additive and the bias completely depends then on the position of the ISM lines with respect to each other and of the photospheric lines. Indeed, the saturated component may hide the non-saturated ones.

\begin{figure}[]
\includegraphics[width=90mm,clip]{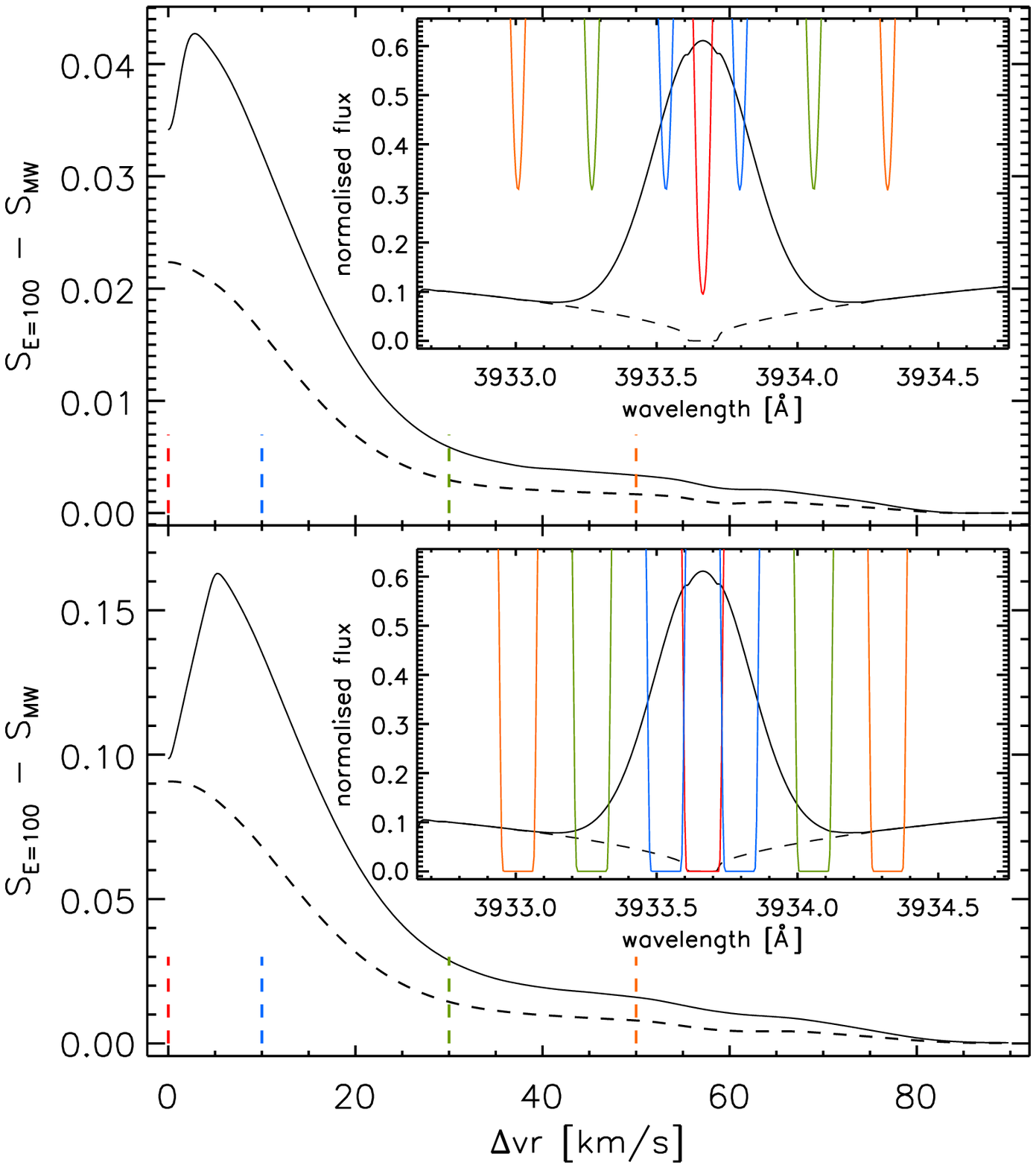}
\caption{Top: difference between the reference $S$ value \So\ calculated without ISM absorption, and \Smw\ obtained considering two non-saturated ($\log N_{\rm CaII}$\,$=$\,11.5) ISM absorption components as a function of velocity separation ($\Delta$vr in \kms) of each ISM feature from the line centre. With increasing $\Delta$vr, the two features move in opposite directions. The black dashed line shows the case of a single ISM absorption component moving redwards of the line centre. The short dashed vertical lines are colour-coded by the absorption features shown in the insert and indicate their position relative to the line centre. The insert shows the photospheric plus chromospheric flux (the photospheric spectrum is shown by a dashed line) in black and, as an example, absorption components lying at a velocity separation of 0.0 (red), $\pm$10.0 (blue), $\pm$30.0 (green), and $\pm$50.0 (orange) \kms\ from the line centre. Bottom: as the top panel, but considering saturated ($\log N_{\rm CaII}$\,$=$\,14.0) ISM absorption lines.}
\label{fig:multiple}
\end{figure}

Figure~\ref{fig:multiple} shows two examples of the difference occurring when considering the presence of one or two ISM absorption components, in the case of saturated (bottom panel) and non-saturated (top panel) lines, as a function of velocity separation of the absorption features from the line centre. The plot indicates for example that in the case of saturated features at the line centre, the presence of multiple components leads to about the same bias as that given by one saturated component.

The increase in the number of free parameters does not allow us to show exhaustive plots for the results of this analysis. Multiple components are common: even for the closest sight lines ($<$100\,pc) the average is 1.7 components per sight line \citep[e.g.][]{redfield2004}. For distant stars we are therefore very likely to encounter the case of multiple ISM absorption lines at different velocities. It is not possible to define averages in terms of number and radial velocity of the ISM components, particularly beyond 100\,pc \citep[e.g.][]{wyman2013}. In most cases, our work can therefore provide only an upper limit (i.e. we assume the presence of one ISM component at the line core) of the effect of ISM absorption. Dedicated simulations, based on the measured radial velocity and strength of the different absorption components, need to be done for a more reliable assessment of the bias on the $S$ index for a particular star.
\section{Effects of ISM absorption on the $S$ index as a function of stellar effective temperature}\label{sec:ISM_general_all}
With decreasing stellar effective temperature, the photospheric flux in the region of the \ion{Ca}{ii}\,H\&K lines decreases. This implies that, with decreasing temperature and at equal chromospheric emission $E$, the $S$ value increases because the emission becomes more prominent compared to that of the nearby continuum (for active late-type stars the $S$ index can be larger than unity). The effect of ISM absorption on \Smw\ is therefore also dependent on the stellar effective temperature.

The right panel of Fig.~\ref{fig:changing_N_VR_Ca2em} shows the differences between \So, calculated for different values of the chromospheric emission $E$, and \Smw\ obtained by varying the ISM parameters for the coolest (\Teff\,=\,4000\,K) star for which we calculated synthetic fluxes. A comparison between the two panels of Fig.~\ref{fig:changing_N_VR_Ca2em} indicates that the stronger the emission, relative to the nearby stellar continuum, the stronger can be the effect of ISM absorption on the derived \Smw.

When comparing the left and right panels of Fig.~\ref{fig:changing_N_VR_Ca2em}, it is important to keep in mind that stars with the same values of $E$, but different effective temperatures, have also different surface chromospheric emissions per cm$^{-2}$. This is because $E$ is the disk-integrated flux at a distance of 1\,AU and stars with different temperatures have different radii. As an example, at equal $E$, there is a difference of a factor of about 4.5 in the surface chromospheric emission per cm$^{-2}$ between the two stars considered in Fig.~\ref{fig:changing_N_VR_Ca2em}.
\section{Discussion}\label{sec:discussion}
\subsection{How relevant is the effect of ISM absorption on activity measurements?}\label{sec:ism}
\begin{figure}[]
\includegraphics[width=90mm,clip]{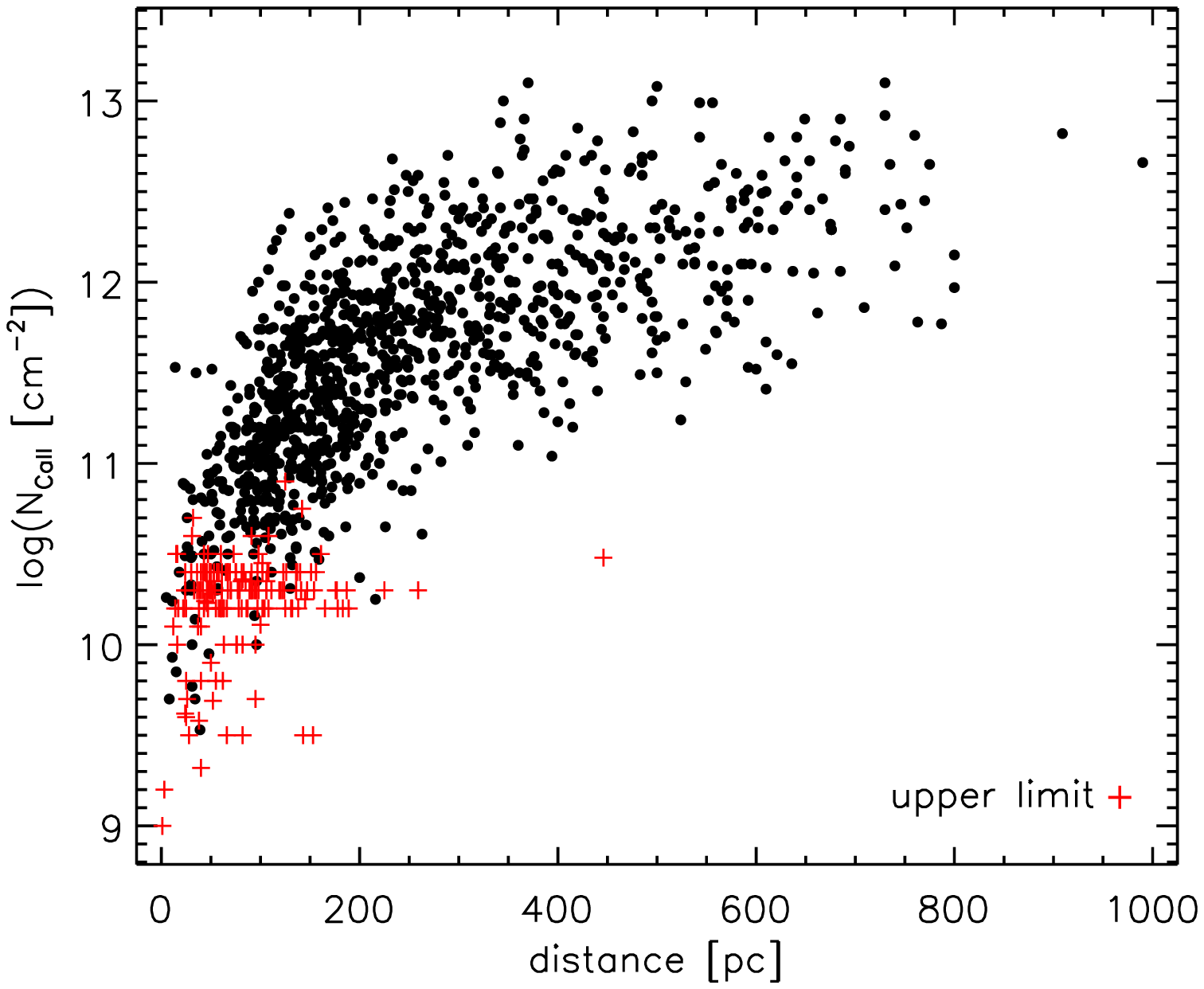}
\includegraphics[width=90mm,clip]{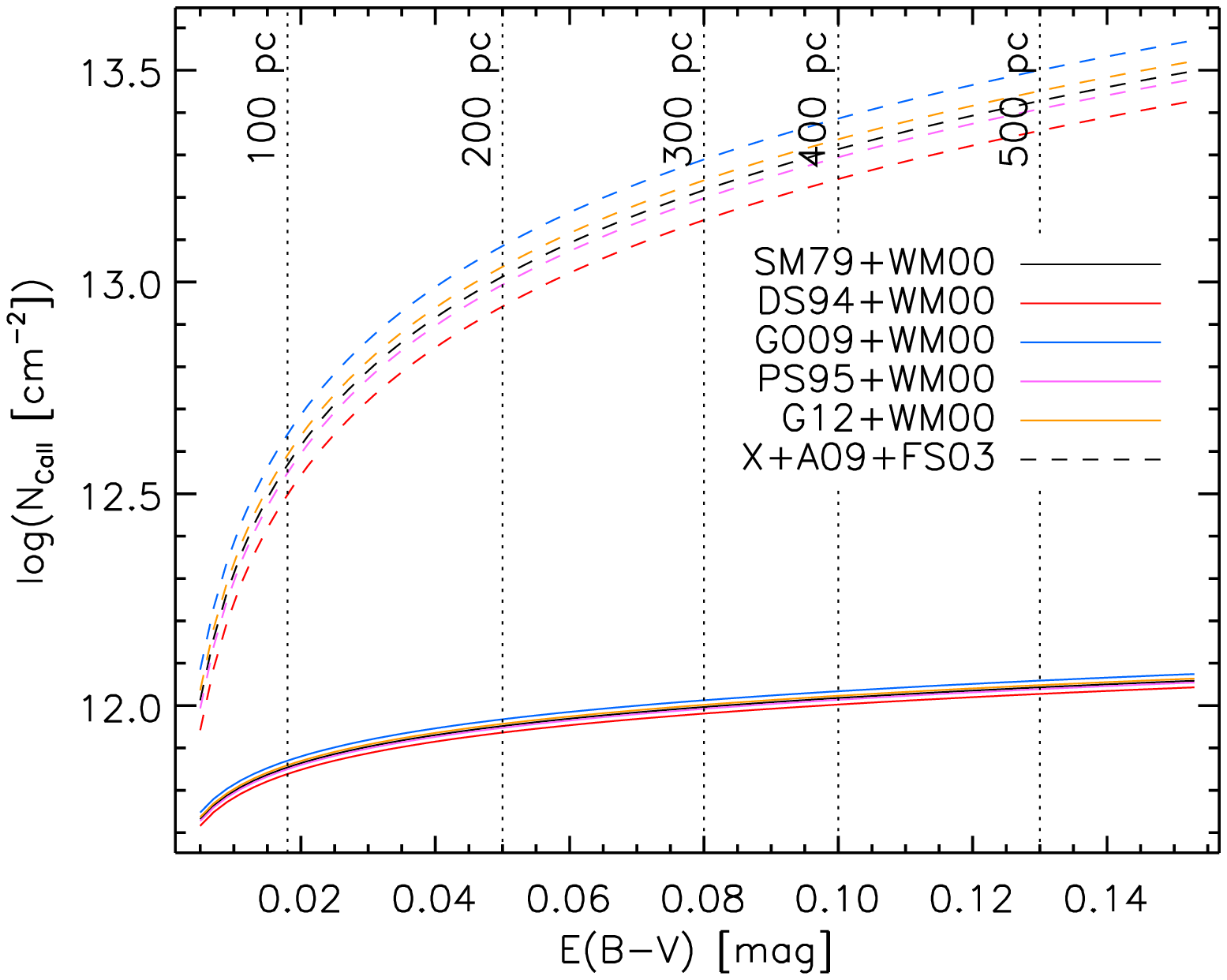}
\caption{Top: measurements of $\log N_{\rm CaII}$ as a function of stellar distance collected by \citet{welsh2010}. Red pluses indicate upper limits. Bottom: $\log N_{\rm CaII}$--$E(B-V)$ correlations obtained by converting first $E(B-V)$ into $\log N_{\rm HI}$, using the relation given by \citet[][SM79; black]{savage1979}, \citet[][DS94; red]{diplas1994}, \citet[][GO09; blue]{guver2009}, \citet[][PS95; purple]{predehl1995}, and \citet[][G12; yellow]{gudennavar2012}, and then $\log N_{\rm HI}$ into $\log N_{\rm CaII}$, using either the $\log N_{\rm HI}$--$\log N_{\rm CaII}$ relation given by \citet[][WM00; solid lines]{wakker2000} or by using the \ion{Ca}{ii} ISM ionisation fraction measured by \citet[][FS03]{frisch2003} and assuming a solar Ca abundance \citep[][A09; dashed lines]{asplund2009}. The ``X'' in the legend stands for any of SM79, DS94, GO09, PS95, or G12. The difference between the solid and dashed lines is caused by the difference in the considered ISM \ion{Ca}{ii} abundance. The vertical dotted lines indicate the average distance at which a given $E(B-V)$ value can be expected following the ISM extinction maps of \citet{amores2005}. This figure justifies the upper boundary of $\log N_{\rm CaII}$\,=\,14 used for the analysis carried out in Sects.~\ref{sec:ISM_general_sun} and \ref{sec:ISM_general_all}.}
\label{fig:Nvsreddening}
\end{figure}

To answer this question, we must first estimate the actual \ion{Ca}{ii} column density at different distances and sight lines. The top panel of Fig.~\ref{fig:Nvsreddening} shows measurements of $\log N_{\rm CaII}$ collected by \citet{welsh2010} for a large number of stars at different distances and sight lines. This plot indicates that on average, for galactic stars more distant that 100--200\,pc, one may expect $\log N_{\rm CaII}$ values of the order of 11.5--12. This conclusion is confirmed, for example, by \citet{wyman2013} who derived \ion{Ca}{ii} column densities from a sample of early-type stars along the direction of the past solar trajectory, obtaining $\log N_{\rm CaII}$\,$>$\,11 (about 12 on average) for each star lying beyond 120\,pc.

As described in Sect.~\ref{sec:correction}, it is practical to have an analytical relation between the ISM \ion{Ca}{ii} column density and the interstellar reddening $E(B-V)$. To this end, we first derived the ISM \ion{H}{i} column density from $E(B-V)$ using the relations given by \citet{savage1979}, \citet{diplas1994}, \citet{guver2009}, \citet{predehl1995}, and \citet{gudennavar2012}. We then converted $\log N_{\rm HI}$ into the \ion{Ca}{ii} column density using either the $\log N_{\rm HI}$--$\log N_{\rm CaII}$ relation given by \citet{wakker2000} or by using the \ion{Ca}{ii} ISM ionisation fraction measured by \citet{frisch2003} and assuming a solar Ca abundance \citep{asplund2009}. The $\log N_{\rm CaII}$--$E(B-V)$ relationships obtained in these ways are shown in the bottom panel of Fig.~\ref{fig:Nvsreddening}. We finally used the ISM extinction maps of \citet{amores2005} to extract the average (across the sky) $E(B-V)$ value as a function of distance and marked it in the bottom panel of Fig.~\ref{fig:Nvsreddening}, up to 500\,pc.

The bottom panel of Fig.~\ref{fig:Nvsreddening} displays a large difference, increasing with $E(B-V)$/distance, between the two sets of lines. This difference is caused by the fact that the work of \citet{frisch2003} focuses on the local ISM (LISM), within about 100\,pc of the Sun, while the work of \citet{wakker2000} is based on measurements of objects, mostly at high galactic latitudes, outside the LISM. This difference affects two properties of Ca in the ISM: the abundance and the ionisation fraction. On the one hand, the adopted solar Ca abundance is most likely too high for the average ISM as the Ca abundance is expected to be 60\%--70\% subsolar, at least within 200\,pc \citep{frisch2003}, and much lower beyond the LISM. In fact, \citet{wakker2000}'s measurements indicate a decreasing \ion{Ca}{ii} abundance with increasing distance. On the other hand, \citet{frisch2011} showed that in the LISM 98.4\% of the Ca is doubly ionised (i.e. \ion{Ca}{iii}), while only 1.55\% is singly ionised (i.e. \ion{Ca}{ii}). Beyond the LISM, it is instead the opposite, with \ion{Ca}{ii} being the dominant species.

The set of solid lines in the bottom panel of Fig.~\ref{fig:Nvsreddening} best matches the observations displayed in the top panel, which is why we consider this set of curves to be more reliable. In addition, the stars for which the activity measurements are most affected by the ISM absorption lie beyond the LISM (see below). We must, however, remember that Fig.~\ref{fig:Nvsreddening} gives only a rough indication of the possible \ion{Ca}{ii} column density at a given $E(B-V)$/distance. For example, because of the higher \ion{Ca}{ii} abundance along the galactic plane \citep{gray2016}, one may find there stars presenting $\log N_{\rm CaII}$ values ranging between 14.1 and 15.1 for $E(B-V)$ values from 0.02 to 0.2\,mag. This warning is to indicate that it is always better to measure the ISM towards the line of sight of the star of interest and to consider the results shown in Fig.~\ref{fig:Nvsreddening} when nothing better is available.

For a Sun-like star, assuming no radial velocity displacement between the star and the ISM absorption, the left panel of Fig.~\ref{fig:changing_N_VR_Ca2em} shows that a \ion{Ca}{ii} ISM column density of $\log N_{\rm CaII}$\,=\,12 would lead to an underestimation of \Smw\ ranging between about 0.005 and 0.04, depending on the stellar chromospheric emission. Figure~\ref{fig:conversion} shows that this would correspond to an underestimation of the \logR\ value of about 0.05--0.1\,dex. We thus conclude that for a Sun-like star beyond 100\,pc, the presence of ISM absorption typically leads to an underestimate of the measured \logR\ values by roughly 0.05--0.1\,dex. The bias caused by the ISM absorption may be even larger for cooler stars.

To find out whether such a bias is really relevant or not, we need to compare it to the typical uncertainties obtained when measuring \Smw\ and \logR. Typical photon noise uncertainties in the measurements of \Smw\ are of the order of 0.002 and are usually negligible compared to other uncertainties connected to the various calibrations necessary to derive the \logR\ value. For example, the typical uncertainty considered when transforming the measured $S$ values into the Mount Wilson system (\Smw) is of the order of about 0.015 in \logR\ \citep{gray2003,jenkins2006,mittag2013}.

These considerations indicate that the magnitude of the bias introduced by the ISM absorption for stars beyond 100\,pc is likely to be comparable, or even larger, than the uncertainties related to the measurement of the \Smw\ and \logR\ values. This conclusion is strengthened by the fact that the maximum root mean square (RMS) scatter of \logR\ measurements for a large number of active and inactive stars is of the order of 0.08 \citep{gomes2014}, which is comparable to the bias introduced by the ISM absorption for stars beyond 100\,pc.
\subsection{Relevance for exoplanet studies}\label{sec:exoplanets}
\begin{figure}[]
\includegraphics[width=90mm,clip]{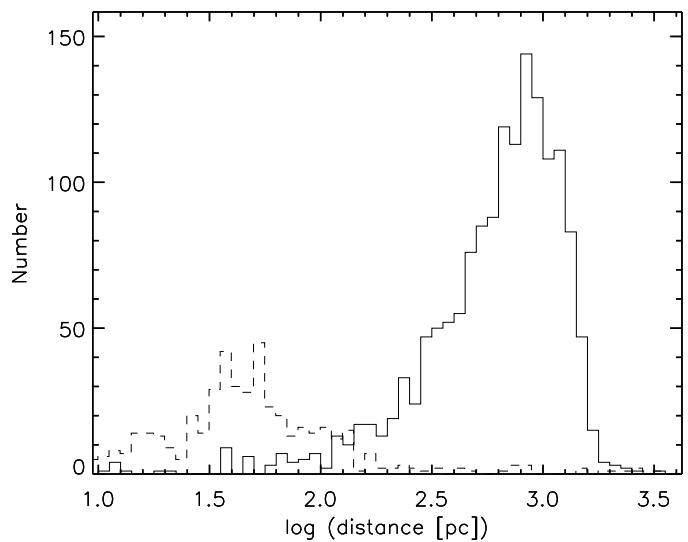}
\caption{Number of transiting (solid line) and non-transiting (dashed line) planets as a function of logarithm of the distance to the solar system, in pc. The distance ranges between about 10\,pc and 3.2\,kpc. The total number of planets is 2167 (1527 transiting and 640 non-transiting). Data downloaded from {\tt http://exoplanets.org/} \citep{han2014} in January 2017.}
\label{fig:distance}
\end{figure}

Classical stellar activity studies concentrate mostly on bright nearby stars, for which the bias due to the ISM absorption is small compared to the other uncertainties. However, the discovery of exoplanets has led to an increasing interest in measuring the activity of more distant stars. Figure~\ref{fig:distance} shows that the majority of known transiting planets lie beyond 100--300\,pc. This is caused by the fact that it is difficult to detect from the ground transiting planets orbiting nearby (hence generally bright) stars and that the Kepler satellite, which discovered the majority of the known transiting planets, concentrated on rather faint and distant stars. We can therefore conclude that the effect of ISM absorption on activity measurements will be relevant for most of the stars currently known to host transiting planets. The most important planets are those transiting bright, nearby stars. For these, detailed follow-up is possible and for these we can get a measurement of the stellar activity which is not likely to be significantly biased by ISM absorption.

Activity measurements are used to guide the detection of planets through radial velocity. In particular, the $S$ index is used to track stellar magnetic cycles, which are then connected to peaks in the periodograms obtained from the radial velocity measurements. On the one hand, within time-scales of tens of years, the ISM absorption can be considered to be stable \citep[e.g.][]{boisse2013}, hence it does not introduce spurious variability that may interfere with planet detections. On the other hand, ISM absorption dims the stellar activity signal, up to the point where it may mask the \ion{Ca}{ii} line core variability, leading to a mis- or non-detection of periodicities connected to stellar rotation and activity cycles, though they may be clearly present in the periodograms obtained from analysing the radial velocity measurements. This would possibly cause misinterpretations of the periodograms, affecting in turn planet detections. However, the careful analysis of the bisectors of the cross-correlation function can usually pin down spurious radial velocity signals related to stellar rotation and activity.

On the other hand, if the ISM absorption is not strong enough to completely mask the detection of stellar magnetic cycles, the ISM absorption has no impact on the derivation of the radial velocity jitter from the activity measurements, at least if one adopts a linear correlation between the radial velocity variation induced by stellar activity and the $S$ index. This is because the slope of the correlation is not modified when subtracting a constant offset from the true $S$ index. The situation would be different if a non-linear correlation is adopted. In this case the subtracted offset matters and the effect could become important for distant stars.

ISM absorption may also affect the relation between stellar activity and planet surface gravity \citep{hartman2010,figueira2014,lanza2014,fossati2015a}. \citet{fossati2015a} analysed this relation concluding that it has a bimodal nature. For their analysis, \citet{fossati2015a} selected planetary systems with stellar magnitude $V$\,$\leq$\,13\,mag and expected colour excess $E(B-V)$\,$\leq$\,0.06\,mag. This was derived from the maps of \citet{amores2005}, with the aim of reducing the systematic effects of ISM absorption on the \logR\ measurements.

Nevertheless, on average more distant stars are more affected by ISM absorption and are also apparently fainter. So planets with large radii are detected in transit around them, while small transiting planets would be undetected for these stars. In addition, on average planets with larger radii have lower gravities, hence one may statistically expect to have a lower \logR\ associated to a star hosting a planet with a lower gravity, in line with the observed correlation. To test this hypothesis we first took the systems considered by \citet{fossati2015a} and looked for a correlation between planet surface gravity and distance to the system. We obtained a Spearman-rank correlation coefficient of 0.035, with a probability of chance occurrence (i.e. no correlation) of 0.843, and a Pearson correlation coefficient of 0.085, indicating that no correlation is found between these parameters. This means that the bias introduced by ISM absorption on the \logR\ measurements should have no effect on the observed stellar activity versus planet surface gravity correlation.

\begin{figure}[]
\includegraphics[width=90mm,clip]{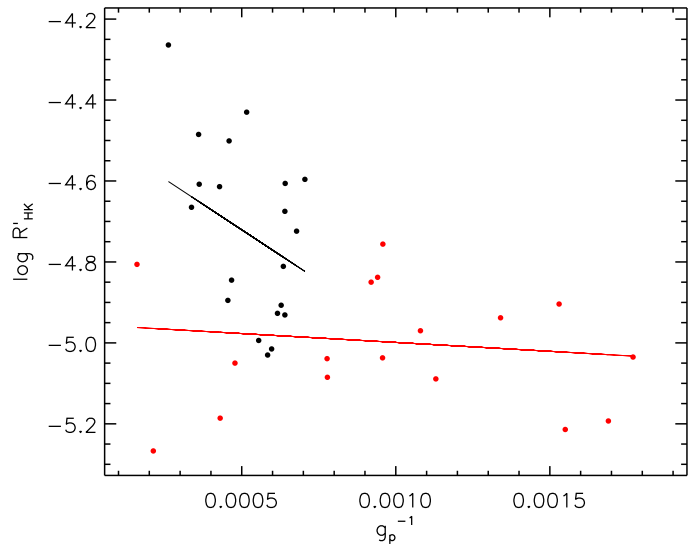}
\caption{Chromospheric emission index \logR\ versus the inverse of the planet gravity g$_p^{-1}$ (in cm$^{-2}$) with the two best-fit regression lines of the mixture model, described in \citet{fossati2015a}, in black and red. The data points assigned to each of the two regressions are plotted with the same colour coding of the corresponding regression line. The high- and low-activity components have respectively an intercept of $-$4.47$\pm$0.14 and $-$4.95$\pm$0.06 and a slope of $-$501$\pm$247 and $-$44$\pm$54 cm\,s$^{-2}$.}
\label{fig:2slopes}
\end{figure}

To strengthen this conclusion, we converted the $E(B-V)$ derived for each star considered by \citet[][see their Sect.~2]{fossati2015a} into $\log N_{\rm CaII}$ using the \ion{H}{i} column density vs $E(B-V)$ relation by \citet{diplas1994} and the $\log N_{\rm HI}$ vs $\log N_{\rm CaII}$ relation given by \citet{wakker2000}. We then used the tools described in Sect.~\ref{sec:input} and the available stellar parameters \citep{fossati2015a} to estimate for each star the maximum correction in the \logR\ value caused by ISM absorption (see Sect.~\ref{sec:correction} for more information). The corrections we applied to the \logR\ values are listed in Table~\ref{tab:corrections}. We finally applied the derived corrections and repeated the statistical analysis thoroughly described in Sect.~3 of \citet{fossati2015a}, using the cluster-weighted model (CWM) proposed in \citet{ingrassia2014}. In particular, a statistical model based on t-student components has been considered. The results of the analysis are shown in Fig.~\ref{fig:2slopes}. We still find that the stellar activity versus planet surface gravity correlation has a bimodal nature, with the modulus of the slope being directly proportional to the intercept, as predicted by the model of \citet{lanza2014}. We conclude therefore that ISM absorption does not significantly affect the stellar activity versus planet surface gravity correlation and hence it does not modify the main results of \citet{fossati2015a}.
\input{stars_complete.tex}
%
\subsection{Correcting activity measurements for the ISM absorption bias}\label{sec:correction}
As demonstrated by the analyses and results presented in Sects.~\ref{sec:ISM_general_sun} and \ref{sec:ISM_general_all}, correcting activity measurements for the ISM absorption is not trivial and it depends on a large number of factors, some of them very difficult, if not impossible, to estimate precisely. For example, to reliably derive the corrections, one would actually need to know {\it a priori} the unbiased stellar chromospheric emission and the ISM parameters (broadening, column density, and radial velocity). The ISM parameters could be obtained for example from extinction maps, such as those of \citet{amores2005}, though this may present rather large uncertainties. This approach may be useful for statistical analyses of large samples. For inactive stars, one could also use the \ion{Na}{i}\,D lines to identify the ISM \ion{Na}{i} column density and radial velocity. The \ion{Na}{i} column density could then be converted into a \ion{Ca}{ii} column density using relations, such as those given by \citet{vallerga1993}, \citet{welty1996}, \citet{welsh2002}, \citet{welsh2003}, \citet{welsh2005}, \citet{welsh2009}, or \citet{wyman2013}. However, the relation between \ion{Na}{i} and \ion{Ca}{ii} is usually inaccurate beyond the LISM (i.e. $\approx$100\,pc), where \ion{Na}{i} and \ion{Ca}{ii} become the dominant ionisation states. The optimal way to measure the ISM parameters would certainly be to directly measure the ISM absorption from a high resolution spectrum of early-type stars lying in the field of view of the target star, if this is a possibility \citep[see e.g.][]{fossati2013,fossati2014}. In this case, however, one could directly correct the spectrum of the target star before measuring the activity.

Inferring the unbiased (i.e. ISM-free) chromospheric emission is problematic for stars located beyond the Local Bubble. Ideally, one could use other activity indicators, possibly not affected by the ISM absorption, such as emission lines of highly ionised elements (e.g. \ion{Si}{iv}) present in the far ultraviolet spectral region. The strength of these lines could then be converted into a \ion{Ca}{ii} emission flux using, for example, the scaling relations of \citet{linsky2013,linsky2014}. \citet{fossati2015b} followed this procedure for the analysis of the WASP-13 system. The major problem with this approach is that it is currently not possible to detect the far ultraviolet spectrum of main-sequence solar-like stars unless they are bright and lie not too far away from us ($\lessapprox$200--250\,pc). This renders the procedure impractical for most of the known transiting planets. This may change though in the next 20-25 years with the next generation of space telescopes with ultraviolet capabilities, such as LUVOIR, for which the larger aperture and higher efficiency of the instruments \citep{france2016} would enable the detection of the far ultraviolet spectra of more distant stars.

Despite the difficulty and possible unreliability of inferring corrections to the activity parameters, we compiled a code that allows the user to roughly estimate the correction in terms of \Smw\ and \logR\ because of ISM absorption. This is the code we adopted to correct the \logR\ values for the stars analysed by \citet{fossati2015a} (Sect.~\ref{sec:exoplanets}). The code is available in both IDL and Python, and can be downloaded at the link {\tt http://geco.oeaw.ac.at/software.html}. The code asks the user to insert (either at terminal or through an input file) the stellar $B-V$ colour or \Teff, the measured \Smw\ or \logR\ values, the colour excess $E(B-V)$ or the ISM $\log N_{\rm CaII}$, and the radial velocity of the ISM absorption feature with respect to the star. The code assumes the presence of one ISM absorption line and uses the library of photospheric fluxes described in Sect.~\ref{sec:input}. The code derives the chromospheric emission flux using the \Smw\ and/or \logR\ values provided by the user (if the \logR\ lies below the basal level of $-$5.1, the code automatically sets \logR\,=\,$-$5.1, which is the minimum possible value without extrinsic absorption). Finally, the code allows the user to choose between two ways of converting the colour excess $E(B-V)$ into $\log N_{\rm CaII}$ following the procedures described in the first paragraph of Sect.~\ref{sec:ism} and adopting the \ion{H}{i} column density versus $E(B-V)$ relation by \citet{diplas1994}. The code assumes infinite spectral resolution, no line broadening from rotation and/or macroturbulence, and a broadening $b$-parameter for the ISM absorption line of 2\,\kms. This last parameter can be modified inside the code. When using the results of this code, the user should be aware of the assumptions and caveats described herein.
\section{Conclusion}\label{sec:conclusion}
Optical and ultraviolet observations have revealed that some stars hosting transiting planets, usually hot Jupiters, present an anomalously low activity, well below the basal level of main-sequence late-type stars of \logR\,=\,$-$5.1. \citet{fossati2015b} concluded that this anomaly may be caused by ISM absorption. Inspired by this result, we study in detail the effect of ISM absorption on activity measurements, concentrating particularly on the $S$ and \logR\ indices. To this end, we model the wavelength region covered by the \ion{Ca}{ii}\,H\&K lines employing photospheric synthetic spectra of stars in the 4000--6400\,K temperature range, to which we add varying amounts of chromospheric emission, simulated as a single Gaussian, and ISM absorption, simulated as a single Voigt profile.

We first considered a Sun-like star with a total chromospheric emission at 1\,AU of 100\,\ergscm, fixed the spectral resolution at 100\,000, assumed a non-rotating star, fixed the \ion{Ca}{ii} ISM column density at $\log N_{\rm CaII}$\,=\,12, and varied the relative velocity between the stellar and ISM spectral features. We found that the bias caused by ISM absorption is significant for relative radial velocities below 30--40\,\kms, in modulus. The bias becomes progressively less important for higher velocities, up to about $\pm$86\,\kms\ where the ISM absorption line starts falling outside of the band used to measure the $S$ index. We then fixed the relative velocity to 0\,\kms\ and varied the \ion{Ca}{ii} column density, keeping all other parameters fixed. We found that the bias exceeds photon noise uncertainties on the $S$ value for \ion{Ca}{ii} logarithmic column densities greater than 10.5. We found also that the bias increases with increasing chromospheric emission.

We explored the effect of varying: relative velocity and \ion{Ca}{ii} column density, spectral resolution, stellar projected rotation velocity, ISM broadening $b$-parameter, and chromospheric emission, one at a time. We found that varying the spectral resolution, stellar projected rotation velocity, and $b$-parameter produces similar, and relatively small, changes on the bias caused by ISM absorption. We conclude that the amount of chromospheric emission plays the greatest role. We found that the bias caused by ISM absorption increases with decreasing stellar effective temperature.

We found for relative radial velocities smaller than 30\,\kms, in modulus, that the bias that ISM absorption causes on the \Smw\ and \logR\ indices exceeds typical photon noise and calibration uncertainties for ISM column densities greater than $\log N_{\rm CaII}$\,$\approx$\,12. Such a column density can be expected for stars beyond 100\,pc. For both active and inactive late-type stars beyond 100\,pc, the bias in activity measurements caused by ISM absorption can be important. The bias increases with increasing activity.

We considered just one ISM absorption line, while observations clearly indicate the presence of multiple ISM clouds, even within very short distances (i.e. $<$\,100\,pc). It is not possible to take this additional level of complication into account in a systematic way. We show that differences from the case of a single line occur when the ISM absorption components do not have the same radial velocity and if one (or more) of the absorption lines is saturated.

Since the ISM absorption differs according to the position of a star in the sky and its effects on activity measurements depend on the number and radial velocity of the ISM components, and not just on the stellar distance, it is not possible to quantify the bias introduced by ISM absorption as a function of distance, expecting that it would be valid across the whole sky. However, our results clearly show that ISM absorption does significantly affect activity measurements. Hence activity studies should either impose a reasonable distance cut-off on their samples (e.g. $\approx$100\,pc) or statistically account for how ISM absorption may affect the results.

ISM absorption should be taken into account by either directly measuring the ISM \ion{Ca}{ii} column density \citep[as done by][for WASP-12]{fossati2013} or by inferring the \ion{Ca}{ii} column density from the extinction (e.g. from fitting of the spectral energy distribution or extinction maps) and calibrations published in the literature. There are many assumptions implicit in the latter method, so it is much more uncertain than the former, though it allows one to infer a rough correction for any star.

We provide a routine to roughly account for ISM absorption when measuring the \Smw\ and \logR\ indices using the latter method. We employ this code to show that the bias caused by ISM absorption does not affect the latest interpretation and results of the stellar activity versus planet surface gravity correlation presented by \citet{fossati2015a}.

Our results could be used to identify whether the anomalously low activity level of certain planet-hosting stars is caused by ISM absorption or by the presence of a translucent circumstellar cloud or torus, as suggested by \citet{haswell2012}. In the latter case, the \logR\ value corrected for the ISM absorption would still lie well below the basal level.
\begin{acknowledgements}
SEM acknowledges financial support from the FEMTECH programme financed by the Austrian Forschungsf\"orderungs-gesellschaft (FFG). DS is supported by an STFC studentship. JSJ acknowledges funding by Fondecyt through grant 1161218, partial support from CATA-Basal (PB06, Conicyt). CH is supported by STFC under grant ST/L000776/1. The authors thank the anonymous referee for the valuable comments. This research has made use of the Exoplanet Orbit Database and the Exoplanet Data Explorer at exoplanets.org.
\end{acknowledgements}
\begin{appendix}
\section{Effect of spectral resolution and stellar rotation on the $S$ values calculated including ISM absorption}
%
\begin{figure*}[h!]
\includegraphics[width=185mm,clip]{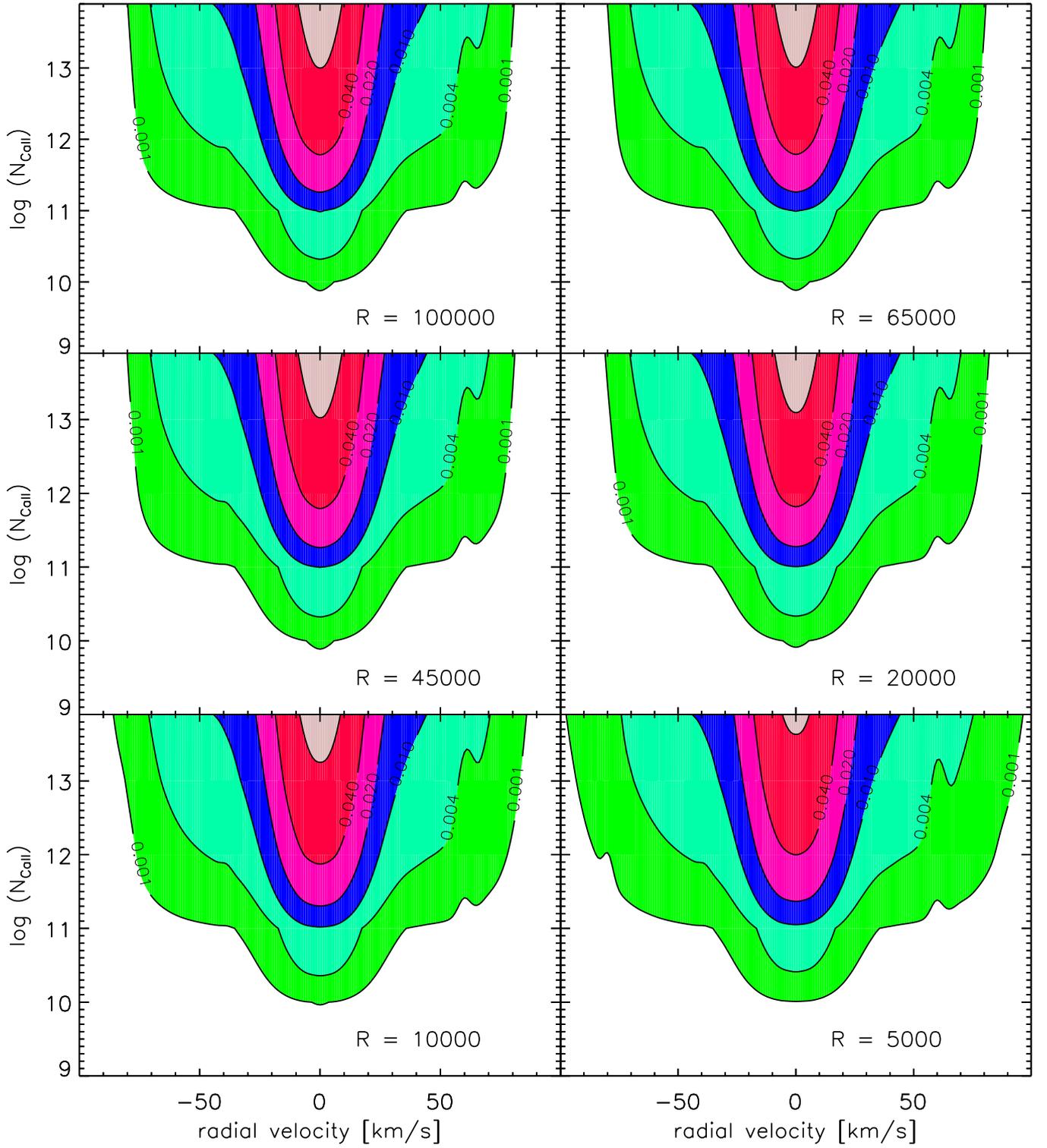}
\caption{Difference between \So\ and the $S$ values obtained by varying the spectral resolution and the ISM parameters (RV and $\log N_{\rm CaII}$), keeping $E$ fixed at 100\,\ergscm. The values along the lines of each contour (0.001, 0.004, 0.01, 0.02, 0.04, 0.08) quantify the \So$-$S difference. The adopted spectral resolution is written in the bottom-right corner of each panel.}
\label{fig:changing_N_VR_R}
\end{figure*}
\begin{figure*}[h!]
\includegraphics[width=185mm,clip]{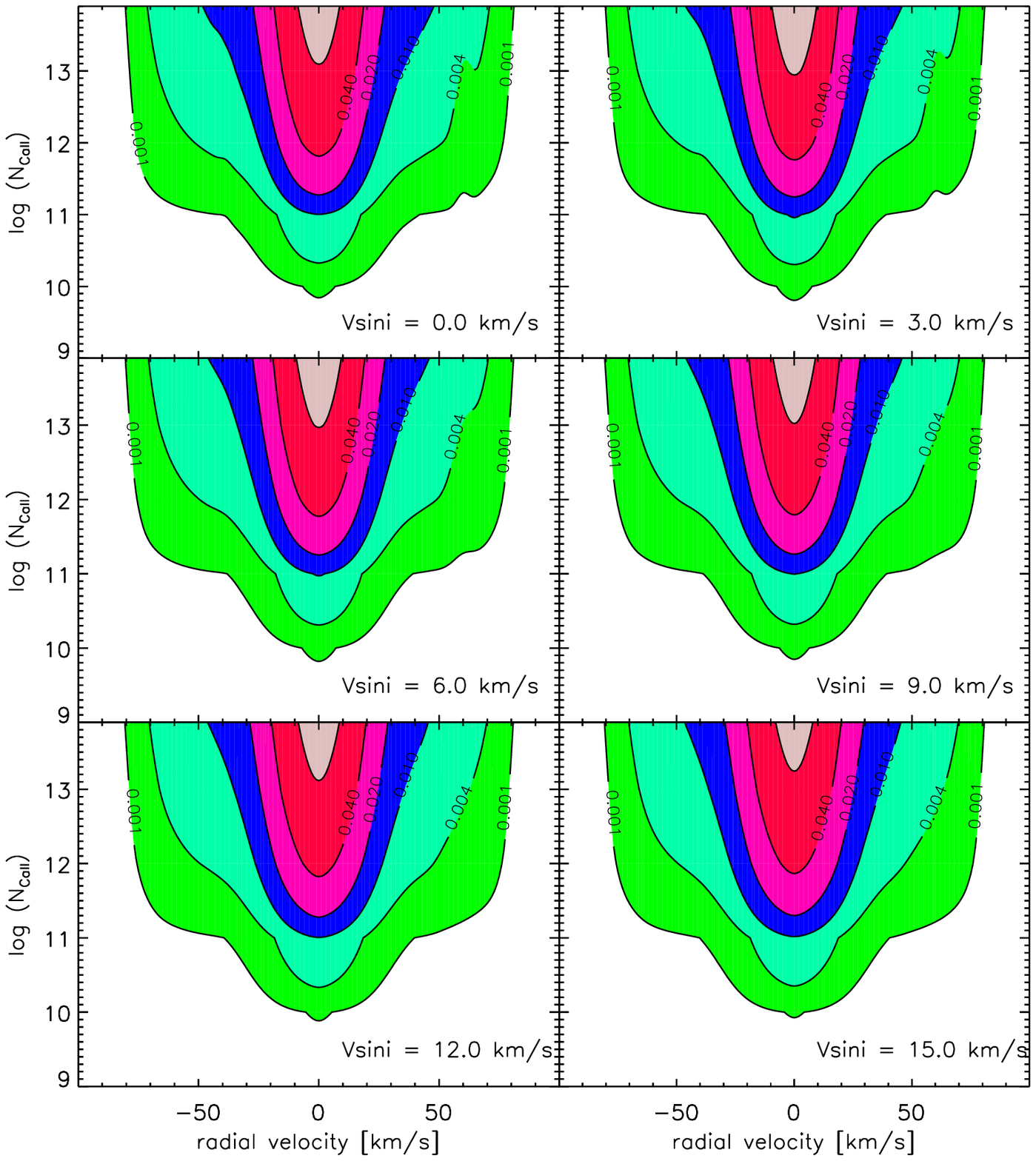}
\caption{Same as Fig.~\ref{fig:changing_N_VR_R}, but varying stellar \vsini.}
\label{fig:changing_N_VR_vsini}
\end{figure*}
%
\end{appendix}
%
%
%
%
%
\end{document}

%% file: stars_complete.tex
\begin{table}[ht]
\caption[ ]{Corrections to the \logR\ values due to ISM absorption for the systems considered in the work of \citet{fossati2015a}. Columns two and three list the correction to the \logR\ values and the corrected \logR\ values, respectively. Columns four and five give the planets' surface gravity (in cgs) and relative uncertainty, as given by \citet{fossati2015a}.}
\label{tab:corrections}
\begin{center}
\begin{tabular}{l|cc|cc}
\hline
\hline
System & $\Delta$\logR & \logR & log\,g$_{\rm p}$ & $\sigma$log\,g$_{\rm p}$ \\
\hline
CoRoT-2		& 0.067 & $-$4.264 & 3.582 & 0.026 \\
HAT-P-1		& 0.046 & $-$4.938 & 2.874 & 0.020 \\
HAT-P-3		& 0.059 & $-$4.845 & 3.331 & 0.056 \\
HAT-P-4		& 0.045 & $-$5.037 & 3.019 & 0.041 \\
HAT-P-5		& 0.046 & $-$5.015 & 3.224 & 0.053 \\
HAT-P-12	& 0.069 & $-$5.035 & 2.753 & 0.030 \\
HAT-P-13	& 0.053 & $-$5.085 & 3.109 & 0.054 \\
HAT-P-16	& 0.056 & $-$4.806 & 3.796 & 0.043 \\
HAT-P-17	& 0.000 & $-$5.039 & 3.110 & 0.028 \\
HAT-P-27	& 0.061 & $-$4.724 & 3.169 & 0.057 \\
HAT-P-31	& 0.045 & $-$5.267 & 3.673 & 0.139 \\
HAT-P-44	& 0.054 & $-$5.193 & 2.773 & 0.070 \\
HD\,149026	& 0.000 & $-$5.030 & 3.233 & 0.449 \\
HD\,189733	& 0.000 & $-$4.501 & 3.338 & 0.056 \\
HD\,209458	& 0.000 & $-$4.970 & 2.968 & 0.015 \\
TrES-1		& 0.063 & $-$4.675 & 3.194 & 0.038 \\
TrES-2		& 0.054 & $-$4.895 & 3.342 & 0.025 \\
TrES-3		& 0.064 & $-$4.485 & 3.444 & 0.058 \\
WASP-2		& 0.060 & $-$4.994 & 3.256 & 0.034 \\
WASP-4		& 0.054 & $-$4.811 & 3.197 & 0.024 \\
WASP-5		& 0.055 & $-$4.665 & 3.472 & 0.045 \\
WASP-11		& 0.067 & $-$4.756 & 3.019 & 0.037 \\
WASP-13		& 0.049 & $-$5.214 & 2.810 & 0.055 \\
WASP-16		& 0.050 & $-$5.050 & 3.320 & 0.063 \\
WASP-19		& 0.064 & $-$4.596 & 3.152 & 0.021 \\
WASP-22		& 0.050 & $-$4.850 & 3.036 & 0.030 \\
WASP-23		& 0.066 & $-$4.614 & 3.368 & 0.061 \\
WASP-26		& 0.049 & $-$4.931 & 3.194 & 0.049 \\
WASP-41		& 0.064 & $-$4.606 & 3.193 & 0.056 \\
WASP-42		& 0.062 & $-$4.838 & 3.026 & 0.052 \\
WASP-48		& 0.046 & $-$5.089 & 2.946 & 0.054 \\
WASP-50		& 0.062 & $-$4.608 & 3.441 & 0.028 \\
WASP-70		& 0.044 & $-$5.186 & 3.366 & 0.098 \\
WASP-84		& 0.000 & $-$4.430 & 3.287 & 0.026 \\
WASP-117	& 0.046 & $-$4.904 & 2.815 & 0.057 \\
XO-1		& 0.051 & $-$4.907 & 3.202 & 0.043 \\
XO-2		& 0.061 & $-$4.927 & 3.210 & 0.029 \\
\hline
\end{tabular}
\end{center}
\end{table}

%% file: Sindex_correction-v6.0.bbl
\begin{thebibliography}{}
\bibitem[Alexander et al.(2016)]{alexander2016}
	Alexander, R.~D., Wynn, G.~A., Mohammed, H., Nichols, J.~D., \& 			Ercolano, B.\ 2016, \mnras, 456, 2766
\bibitem[Am{\^o}res \& L{\'e}pine(2005)]{amores2005}
	Am{\^o}res, E.~B., \& L{\'e}pine, J.~R.~D.\ 2005, \aj, 130, 659
\bibitem[Asplund et al.(2009)]{asplund2009}
	Asplund, M., Grevesse, N., Sauval, A.~J., \& Scott, P.\ 2009, \araa, 47, 	481
\bibitem[Baliunas et al.(1995)]{baliunas1995}
	Baliunas, S.~L., Donahue, R.~A., Soon, W.~H., et al.\ 1995, \apj, 438, 			269
\bibitem[Barnes et al.(2016)]{barnes2016}
	Barnes, J.~R., Haswell, C.~A., Staab, D., \& Anglada-Escud{\'e}, G.\ 			2016, \mnras, 462, 1012
\bibitem[Bisikalo et al.(2013)]{bisikalo2013}
	Bisikalo, D., Kaygorodov, P., Ionov, D., et al.\ 2013, \apj, 764, 19
\bibitem[Boiss{\'e} et al.(2013)]{boisse2013}
	Boiss{\'e}, P., Federman, S.~R., Pineau des For{\^e}ts, G., \& Ritchey, 		A.~M.\ 2013, \aap, 559, A131
\bibitem[Bourrier et al.(2016)]{bourrier2016}
	Bourrier, V., Ehrenreich, D., King, G., et al.\ 2016, \aap, 597, A26
\bibitem[Canto Martins et al.(2011)]{canto2011}
	Canto Martins, B.~L., Das Chagas, M.~L., Alves, S., et al.\ 2011, \aap, 		530, A73
\bibitem[Carroll-Nellenback et al.(2016)]{carroll2016}
	Carroll-Nellenback, J., Frank, A., Liu, B., et al.\ 2017, \mnras, 466, 			2458
\bibitem[Cauley et al.(2016)]{cauley2016}
	Cauley, P.~W., Redfield, S., Jensen, A.~G., \& Barman, T.\ 2016, \aj, 			152, 20
\bibitem[Cayrel de Strobel(1996)]{cayrel1996}
	Cayrel de Strobel, G.\ 1996, \aapr, 7, 243
\bibitem[Diplas \& Savage(1994)]{diplas1994}
	Diplas, A. \& Savage, B.~D. 1994, \apj, 427, 274
\bibitem[Ehrenreich et al.(2012)]{ehrenreich2012}
	Ehrenreich, D., Bourrier, V., Bonfils, X., et al.\ 2012, \aap, 547, A18
\bibitem[Ehrenreich et al.(2015)]{ehrenreich2015}
	Ehrenreich, D., Bourrier, V., Wheatley, P.~J., et al.\ 2015, \nat, 522, 		459
\bibitem[Erkaev et al.(2016)]{erkaev2016}
	Erkaev, N.~V., Lammer, H., Odert, P., et al.\ 2016, \mnras, 460, 1300 \bibitem[Figueira et al.(2014)]{figueira2014}
	Figueira, P., Oshagh, M., Adibekyan, V.~Z., \& Santos, N.~C.\ 2014, 			\aap, 572, A51
\bibitem[Fossati et al.(2010a)]{fossati2010a}
	Fossati, L., Haswell, C.~A., Froning, C.~S., et al.\ 2010a, \apjl, 714, 		L222
\bibitem[Fossati et al.(2010b)]{fossati2010b}
	Fossati, L., Bagnulo, S., Elmasli, A., et al.\ 2010b, \apj, 720, 872
\bibitem[Fossati et al.(2013)]{fossati2013}
	Fossati, L., Ayres, T.~R., Haswell, C.~A., et al.\ 2013, \apjl, 766, L20
\bibitem[Fossati et al.(2014)]{fossati2014}
	Fossati, L., Ayres, T.~R., Haswell, C.~A., et al.\ 2014, \apss, 354, 21
\bibitem[Fossati et al.(2015a)]{fossati2015a}
	Fossati, L., Ingrassia, S., \& Lanza, A.~F.\ 2015a, \apjl, 812, L35 \bibitem[Fossati et al.(2015b)]{fossati2015b}
	Fossati, L., France, K., Koskinen, T., et al.\ 2015b, \apj, 815, 118
\bibitem[France et al.(2016)]{france2016}
	France, K., Fleming, B., \& Hoadley, K.\ 2016, \procspie, 9905, 990506
\bibitem[Frisch \& Slavin(2003)]{frisch2003}
	Frisch, P.~C., \& Slavin, J.~D.\ 2003, \apj, 594, 844
\bibitem[Frisch et al.(2011)]{frisch2011} 
	Frisch, P.~C., Redfield, S., \& Slavin, J.~D.\ 2011, \araa, 49, 237 
\bibitem[Garc{\'{\i}}a Mu{\~n}oz(2007)]{gm2007}
	Garc{\'{\i}}a Mu{\~n}oz, A.\ 2007, \planss, 55, 1426
\bibitem[Gomes da Silva et al.(2014)]{gomes2014}
	Gomes da Silva, J., Santos, N.~C., Boisse, I., Dumusque, X., \& Lovis, 			C.\ 2014, \aap, 566, A66
\bibitem[Gray et al.(2003)]{gray2003}
	Gray, R.~O., Corbally, C.~J., Garrison, R.~F., McFadden, M.~T., \& 			Robinson, P.~E.\ 2003, \aj, 126, 2048
\bibitem[Gray(2005)]{gray2005}
	Gray, D.~F.\ 2005, ``The Observation and Analysis of Stellar 				Photospheres'', 3rd Edition, by D.F.~Gray.~ISBN	0521851866, UK: 			Cambridge University Press, 2005
\bibitem[Gray \& Scannapieco(2016)]{gray2016}
	Gray, W.~J., \& Scannapieco, E.\ 2016, \apj, 818, 198
\bibitem[Gudennavar et al.(2012)]{gudennavar2012}
	Gudennavar, S.~B., Bubbly, S.~G., Preethi, K. \& Murthy, J. 2012, \apjs, 	199, 8
\bibitem[G{\"u}ver \& {\"O}zel(2009)]{guver2009}
	G{\"u}ver, T., \& {\"O}zel, F.\ 2009, \mnras, 400, 2050
\bibitem[Hall et al.(2007)]{hall2007}
	Hall, J.~C., Lockwood, G.~W., \& Skiff, B.~A.\ 2007, \aj, 133, 862
\bibitem[Han et al.(2014)]{han2014} 
	Han, E., Wang, S.~X., Wright, J.~T., et al.\ 2014, \pasp, 126, 827 
\bibitem[Hartman(2010)]{hartman2010}
	Hartman, J.~D.\ 2010, \apjl, 717, L138
\bibitem[Haswell et al.(2012)]{haswell2012}
	Haswell, C.~A., Fossati, L., Ayres, T., et al.\ 2012, \apj, 760, 79
\bibitem[Haywood et al.(2016)]{haywood2016}
	Haywood, R.~D., Collier Cameron, A., Unruh, Y.~C., et al.\ 2016, \mnras, 	457, 3637
\bibitem[Hebb et al.(2009)]{hebb2009}
	Hebb, L., Collier-Cameron, A., Loeillet, B., et al.\ 2009, \apj, 693, 			1920
\bibitem[Ingrassia et al.(2014)]{ingrassia2014}
	Ingrassia, S., Minotti, S.~C., Punzo, A.\ 2014, Computational Statistics 	\& Data Analysis, 71, 159
\bibitem[Jenkins et al.(2006)]{jenkins2006}
	Jenkins, J.~S., Jones, H.~R.~A., Tinney, C.~G., et al.\ 2006, \mnras, 			372, 163
\bibitem[Jenkins et al.(2008)]{jenkins2008}
	Jenkins, J.~S., Jones, H.~R.~A., Pavlenko, Y., et al.\ 2008, \aap, 485, 		571
\bibitem[Jenkins et al.(2011)]{jenkins2011}
	Jenkins, J.~S., Murgas, F., Rojo, P., et al.\ 2011, \aap, 531, A8
\bibitem[Jensen et al.(2012)]{jensen2012}
	Jensen, A.~G., Redfield, S., Endl, M., et al.\ 2012, \apj, 751, 86
\bibitem[Johnson et al.(2015)]{johson2015}
	Johnson, M.~C., Redfield, S., \& Jensen, A.~G.\ 2015, \apj, 807, 162
\bibitem[Kislyakova et al.(2016)]{kislyakova2016}
	Kislyakova, K.~G., Pilat-Lohinger, E., Funk, B., et al.\ 2016, \mnras, 			461, 988
\bibitem[Knutson et al.(2010)]{knutson2010}
	Knutson, H.~A., Howard, A.~W., \& Isaacson, H.\ 2010, \apj, 720, 1569
\bibitem[Koskinen et al.(2014)]{koskinen2014}
	Koskinen, T.~T., Lavvas, P., Harris, M.~J., \& Yelle, R.~V.\ 2014, 			Philosophical Transactions of the Royal Society of London Series A, 372, 	20130089
\bibitem[Kulow et al.(2014)]{kulow2014}
	Kulow, J.~R., France, K., Linsky, J., \& Loyd, R.~O.~P.\ 2014, \apj, 			786, 132
\bibitem[Lai et al.(2010)]{lai2010}
	Lai, D., Helling, C., \& van den Heuvel, E.~P.~J.\ 2010, \apj, 721, 923
\bibitem[Lammer et al.(2003)]{lammer2003}
	Lammer, H., Selsis, F., Ribas, I., et al.\ 2003, \apjl, 598, L121
\bibitem[Lanza(2014)]{lanza2014}
	Lanza, A.~F.\ 2014, \aap, 572, L6
\bibitem[Lecavelier des Etangs et al.(2004)]{lecav2004}
	Lecavelier des Etangs, A., Vidal-Madjar, A., McConnell, J.~C., \& 			H{\'e}brard, G.\ 2004, \aap, 418, L1
\bibitem[Lecavelier des Etangs et al.(2012)]{lecavelier2012}
	Lecavelier des Etangs, A., Bourrier, V., Wheatley, P.~J., et al.\ 2012, 		\aap, 543, L4
\bibitem[Linsky et al.(2010)]{linsky2010}
	Linsky, J.~L., Yang, H., France, K., et al.\ 2010, \apj, 717, 1291
\bibitem[Linsky et al.(2013)]{linsky2013}
	Linsky, J.~L., France, K., \& Ayres, T.\ 2013, \apj, 766, 69
\bibitem[Linsky et al.(2014)]{linsky2014}
	Linsky, J.~L., Fontenla, J., \& France, K.\ 2014, \apj, 780, 61
\bibitem[Llama et al.(2011)]{llama2011}
	Llama, J., Wood, K., Jardine, M., et al.\ 2011, \mnras, 416, L41 \bibitem[Lovis et al.(2011)]{lovis2011}
	Lovis, C., Dumusque, X., Santos, N.~C., et al.\ 2011, arXiv:1107.5325
\bibitem[Mamajek \& Hillenbrand(2008)]{mamajek2008}
	Mamajek, E.~E. \& Hillenbrand, L.~A. 2008, \apj, 687, 1264
\bibitem[Mayor \& Queloz(1995)]{mayor1995}
	Mayor, M., \& Queloz, D.\ 1995, \nat, 378, 355
\bibitem[Middelkoop(1982)]{middelkoop1982}
	Middelkoop, F.\ 1982, \aap, 107, 31
\bibitem[Miller et al.(2015)]{miller2015}
	Miller, B.~P., Gallo, E., Wright, J.~T., \& Pearson, E.~G.\ 2015, \apj, 		799, 163
\bibitem[Mittag et al.(2013)]{mittag2013}
	Mittag, M., Schmitt, J.~H.~M.~M., \& Schr{\"o}der, K.-P.\ 2013, \aap, 			549, A117
\bibitem[Murray-Clay et al.(2009)]{mc2009}
	Murray-Clay, R.~A., Chiang, E.~I., \& Murray, N.\ 2009, \apj, 693, 23
\bibitem[Nichols et al.(2015)]{nichols2015}
	Nichols, J.~D., Wynn, G.~A., Goad, M., et al.\ 2015, \apj, 803, 9
\bibitem[Noyes et al.(1984)]{noyes1984}
	Noyes, R.~W., Weiss, N.~O., \& Vaughan, A.~H.\ 1984, \apj, 287, 769
\bibitem[Oranje(1983)]{oranje1983}
	Oranje, B.~J.\ 1983, \aap, 122, 88
\bibitem[Owen \& Wu(2016)]{owen2016}
	Owen, J.~E., \& Wu, Y.\ 2016, \apj, 817, 107
\bibitem[Predehl \& Schmitt(1995)]{predehl1995}
	Predehl, P., \& Schmitt, J.~H.~M.~M.\ 1995, \aap, 293, 889
\bibitem[Redfield \& Linsky(2002)]{redfield2002}
	Redfield, S., \& Linsky, J.~L.\ 2002, \apjs, 139, 439
\bibitem[Redfield \& Linsky(2004)]{redfield2004}
	Redfield, S., \& Linsky, J.~L.\ 2004, \apj, 613, 1004
\bibitem[Rutten(1984)]{rutten1984}
	Rutten, R.~G.~M.\ 1984, \aap, 130, 353
\bibitem[Salz et al.(2015)]{salz2015}
	Salz, M., Banerjee, R., Mignone, A., et al.\ 2015, \aap, 576, A21 \bibitem[Salz et al.(2016)]{salz2016}
	Salz, M., Czesla, S., Schneider, P.~C., \& Schmitt, J.~H.~M.~M.\ 2016, 			\aap, 586, A75
\bibitem[Savage \& Mathis(1979)]{savage1979}
	Savage, B.~D. \& Mathis, J.~S. 1979, \araa, 17, 73
\bibitem[Shulyak et al.(2004)]{llm}
	Shulyak, D., Tsymbal, V., Ryabchikova, T., St{\"u}tz, C., \& Weiss, 			W.~W.\ 2004, \aap, 428, 993
\bibitem[Skillen et al.(2009)]{skillen2009}
	Skillen, I., Pollacco, D., Collier Cameron, A., et al.\ 2009, \aap, 502, 	391
\bibitem[Staab et al.(2017)]{staab2016}
	Staab, D., Haswell, C.~A., Smith, G., et al.\ 2017, \mnras, 466, 738
\bibitem[Vallerga et al.(1993)]{vallerga1993}
	Vallerga, J.~V., Vedder, P.~W., Craig, N., \& Welsh, B.~Y.\ 1993, \apj, 		411, 729
\bibitem[Vidal-Madjar et al.(2003)]{vidal-madjar2003}
	Vidal-Madjar, A., Lecavelier des Etangs, A., D{\'e}sert, J.-M., et al.\ 		2003, \nat, 422, 143
\bibitem[Vidal-Madjar et al.(2004)]{vidal-madjar2004}
	Vidal-Madjar, A., D\'esert, J.-M., Lecavelier des Etangs, A., et al.\ 			2004, \apj, 604, L69
\bibitem[Vidotto et al.(2010)]{vidotto2010}
	Vidotto, A.~A., Jardine, M., \& Helling, C.\ 2010, \apjl, 722, L168
\bibitem[Wakker \& Mathis(2000)]{wakker2000}
	Wakker, B.~P., \& Mathis, J.~S.\ 2000, \apjl, 544, L107
\bibitem[Welsh et al.(2002)]{welsh2002}
	Welsh, B.~Y., Rachford, B.~L., \& Tumlinson, J.\ 2002, \aap, 381, 566
\bibitem[Welsh \& Sallmen(2003)]{welsh2003}
	Welsh, B.~Y., \& Sallmen, S.\ 2003, \aap, 408, 545
\bibitem[Welsh et al.(2005)]{welsh2005}
	Welsh, B.~Y., Sallmen, S., \& Jelinsky, S.\ 2005, \aap, 440, 547
\bibitem[Welsh et al.(2009)]{welsh2009}
	Welsh, B.~Y., Wheatley, J., \& Lallement, R.\ 2009, \pasp, 121, 606
\bibitem[Welsh et al.(2010)]{welsh2010} 
	Welsh, B.~Y., Lallement, R., Vergely, J.-L., \& Raimond, S.\ 2010, \aap, 	510, A54 
\bibitem[Welty et al.(1996)]{welty1996}
	Welty, D.~E., Morton, D.~C., \& Hobbs, L.~M.\ 1996, \apjs, 106, 533
\bibitem[Wright(2004)]{wright2004}
	Wright, J.~T.\ 2004, \aj, 128, 1273
\bibitem[Wyman \& Redfield(2013)]{wyman2013}
	Wyman, K., \& Redfield, S.\ 2013, \apj, 773, 96
\bibitem[Yelle(2004)]{yelle2004}
	Yelle, R.~V.\ 2004, \icarus, 170, 167
\end{thebibliography}
